\documentclass[12pt, draftclsnofoot, onecolumn]{IEEEtran}
\usepackage[T1]{fontenc}

\ifCLASSINFOpdf
\else
\fi
\usepackage{amsmath}
\usepackage{amsmath,graphicx}
\usepackage{amssymb,amsfonts}                           
\usepackage{amsthm}
\usepackage{color}
\usepackage{multirow}
\usepackage{mathrsfs}
\usepackage{balance}

\usepackage[tight,footnotesize]{subfigure}

\newtheorem{proposition}{Proposition}
\theoremstyle{definition}
\newtheorem{definition}{Definition}
\newtheorem{lemma}{Lemma}
\newtheorem{corollary}{Corollary}
\theoremstyle{remark}
\newtheorem{remark}{Remark}

\begin{document}
%
\title{Frequency-packed Faster-than-Nyquist Signaling via Symbol-level Precoding for Multi-user MISO Redundant Transmissions}
%
%
%

\author{Wallace A. Martins,~\IEEEmembership{Senior Member,~IEEE,} Symeon Chatzinotas,~\IEEEmembership{Senior Member,~IEEE,} and Bj\"orn Ottersten,~\IEEEmembership{Fellow,~IEEE}
\thanks{The authors are with the Interdisciplinary Centre for Security Reliability and Trust (SnT), University of Luxembourg, Luxembourg. E-mails: {\tt \{wallace.alvesmartins,symeon.chatzinotas,bjorn.ottersten\}@uni.lu}.}
\thanks{This research was funded in whole by the Luxembourg National
Research Fund (FNR) in the frameworks of the FNR-UKRI project ``CI-PHY: Exploiting interference for physical layer security in 5G networks'' (Grant no. FNR/11607830) and of the FNR-CORE project ``RISOTTI: Reconfigurable intelligent surfaces for smart cities'' (Grant no. FNR/14773976). For the purpose of open access,
the authors have applied a Creative Commons Attribution 4.0 International
(CC BY 4.0) license to any Author Accepted Manuscript version arising from
this submission.}}

\markboth{AUTHORS' DRAFT, JULY 2021}%
{Shell \MakeLowercase{\textit{et al.}}: Bare Demo of IEEEtran.cls for IEEE Communications Society Journals}
%



\maketitle

\begin{abstract}
This work addresses the issue of interference generated by co-channel users in downlink multi-antenna multicarrier systems with frequency-packed \emph{faster-than-Nyquist} (FTN) signaling. The resulting interference stems from an aggressive strategy for enhancing the throughput via frequency reuse across different users and the squeezing of signals in the time-frequency plane beyond the Nyquist limit. The error-free spectral efficiency is proved to be increasing with the frequency packing and FTN acceleration factors. The lower bound for the FTN sampling period that guarantees information losslesness is derived as a function of the transmitting-filter roll-off factor, the frequency-packing factor, and the number of subcarriers. Space-time-frequency \emph{symbol-level precoders} (SLPs) that trade off constructive and destructive \emph{interblock interference} (IBI) at the single-antenna user terminals are proposed. Redundant elements are added as guard interval to cope with vestigial destructive IBI effects. The proposals can handle channels with delay spread longer than the multicarrier-symbol duration. The receiver architecture is simple, for it does not require digital multicarrier demodulation.  Simulations indicate that the proposed SLP outperforms zero-forcing precoding and achieves a target balance between spectral and energy efficiencies by controlling the amount of added redundancy from zero (full IBI) to half (destructive IBI-free) the group delay of the equivalent channel.
\end{abstract}


%
\IEEEpeerreviewmaketitle

%
%
%
%

\section{Introduction}
\label{sec:intro}

The fundamentally limited physical resources of wireless communication systems are the wireless spectrum and transmit power. The exact resource which plays the main role in a specific system design depends largely on the application goals and constraints. In the context of \emph{multi-user} (MU) \emph{multiple-input single-output} (MISO)  downlink transmissions, targeting higher data rates is the most common trend, which requires an efficient usage of the available wireless spectrum. From a physical-layer viewpoint, the aggressive frequency reuse across different users as well as the packing of the transmitted signals in the time-frequency plane~\cite{Rusek2005,Anderson2013} are amongst the most promising forms of increasing spectral efficiency. 

When a multi-antenna base station serves several user terminals, the capacity gains stemming from full frequency-reuse downlink transmissions are greatly affected by the underlying \emph{multi-user interference} (MUI). \emph{Symbol-level precoding} (SLP)~\cite{Masouros2009,Masouros2010} encompasses a set of techniques that can benefit from the otherwise harmful MUI effects by shaping the transmitted waveforms so as to induce constructive interference at the user terminals~\cite{Masouros2011,Alodeh2015,Alodeh2015b,Masouros2015}. SLP is a non-linear technique that employs \emph{channel-state information} (CSI) along with users' data to form the precoder. Several SLP schemes exploiting different properties of the communication environment have been proposed~\cite{Alodeh2016,Alodeh2017b,Spano2018a,Choi2019}; the reader is referred to~\cite{Alodeh2018,Ottersten2019} for further details on SLP. 

Besides aggressive frequency reuse, \emph{faster-than-Nyquist} (FTN) signaling~\cite{Mazo1975} has recently been considered as a viable alternative for enhancing spectral efficiency by accelerating the transmission of symbols beyond the Nyquist limit~\cite{Liveris2003,Kim2016}. Although most FTN schemes focus on compensating the introduced \emph{intersymbol interference} (ISI) at the receiver end~\cite{Anderson2013}, some recent works tackle the ISI using precoding  techniques~\cite{Alodeh2017a,Alodeh2017c,Xu2019,Li2019,Jana2019,Martins2020}. When \emph{multicarrier} (MC) systems are employed, one can squeeze the signals in both time and frequency domains~\cite{Rusek2005} via frequency-packed FTN signaling, and very few works have exploited this fact along with precoding designs~\cite{Xu2019,Li2019,Jana2019}. 

It is worth noting that MC-modulated signals are commonly used in broadband communications to overcome the challenges imposed by  frequency-selective channels. MC systems usually employ variations of the well-known \emph{orthogonal frequency-division multiplexing} (OFDM)~\cite{Diniz2012}. Although most OFDM-based schemes focus on compensating ISI at the receiver end, it is possible to adapt well-known precoders, such as \emph{maximum ratio transmitter} (MRT)~\cite{Lo1999} and \emph{zero-forcing} (ZF)~\cite{Nossek2005}, to the multicarrier case, considering the equivalent model of parallel flat-fading channels obtained from a convenient introduction of guard intervals. In this case, the guard-interval length is larger than the equivalent channel order.

Conventional precoders usually handle MUI only in the spatial domain. This also applies to OFDM systems thanks to their equivalent parallel flat-fading model. When such a model is not suitable, the users' data, which are split into  subcarriers (frequency domain) and are  transmitted using an antenna array (spatial domain), may experience both spatial and \emph{intercarrier interference} (ICI), besides ISI stemming from FTN accelerated transmissions. Space-time-frequency precoding is, therefore, called for. 
In fact, the design of space-time-frequency SLP techniques that  induce constructive interference at the receivers is an open problem as of yet. A key aspect in such an open problem is to address how these schemes can cope with the \emph{interblock interference} (IBI) inherent in frequency-selective channels.

In this work, we address this open problem by proposing a new MU-MISO system model that tackles the  frequency-selectivity-related IBI, ISI, and ICI effects using space-time-frequency SLPs. We adapt to the non-linear setup some ideas from reduced-redundancy linear transceivers~\cite{Scaglione1999,Lin2002,Martins2009,Martins2010a,Martins2011,Martins2012a,Martins2012b}, and design redundant space-time-frequency SLPs that minimize the total transmit power while allowing for the trade-off between constructive and destructive IBI effects. The amount of added redundancy may vary from zero (full IBI) to half (destructive IBI-free) the group delay of the equivalent channel model. Thus, in addition to saving bandwidth as compared to conventional OFDM-based systems (since the group delay is usually much smaller than the channel order, and we use at most half the group delay as guard-interval length), the proposed precoders also simplify the receiver architecture by relieving it of performing \emph{discrete Fourier transform} (DFT) computations, as it will be further detailed. We also characterize mathematically the behavior of the error-free spectral efficiency (i.e., the achievable data rate normalized by the bandwidth) showing its monotonicity with respect to the sampling time and frequency-packing factor. We derive the minimum admissable sampling time that allows for information losslessness~\cite{Martins2020} as a function of the frequency-packing factor, the number of subcarriers, and the roll-off factor of the transmiting/receiving filters. We also provide the relations among important variables, like the the minimum guard-interval length that enables destructive IBI-free transmissions as a function of the group delay of the effective channel, as well as the exact number of backward and forward IBI-related blocks affecting the signal reconstruction at the receiver end.

The paper is organized as follows. An MU-MISO system model of a multicarrier linear ZF precoder is described in Section~\ref{sec:zf-mc}. 
A set of new results along with detailed discussions regarding frequency-packed FTN signaling are provided in Secion~\ref{sec:ftn}. 
A new MU-MISO multicarrier system is proposed in Section~\ref{sec:slp-mc} along with the non-linear space-time-frequency SLP. The performance of the proposed precoders is assessed through numerical experiments in Section~\ref{sec:res}. The concluding remarks are in Section~\ref{sec:conc}.

\emph{Notation}: 
Scalars are denoted by italic letters, whereas vectors and matrices 
are denoted by boldface letters (lowercase for vectors and uppercase for matrices). 
Calligraphic letters denote sets. Discrete-time signals are expressed with brackets and continuous-time signals with parentheses; $\delta[n]$ is the Kronecker discrete-time pulse, whereas $\delta(t)$ is the Dirac continuous-time impulse. The Fourier transform of $f(t)$ is denoted as $F({\rm j}\omega)$. The symbols $\triangleq$ and $\ast$ denote definition assignment and linear convolution, respectively, whereas $\otimes$ and $\odot$ denote Kronecker and Hadamard products, respectively. 
The notations $(\cdot)^\intercal$ and $(\cdot)^{\rm H}$ stand for transpose and Hermitian transpose operations on $(\cdot)$, respectively. Given a real number $x$, $\left\lfloor x \right\rfloor$ and $\left\lceil x \right\rceil$ respectively stand for the largest integer smaller than or equal to $x$ and the smallest integer greater than or equal to $x$.

\section{Multicarrier  Precoding Preliminaries}\label{sec:zf-mc}

 \begin{figure*}[!t]
\includegraphics[width=\linewidth]{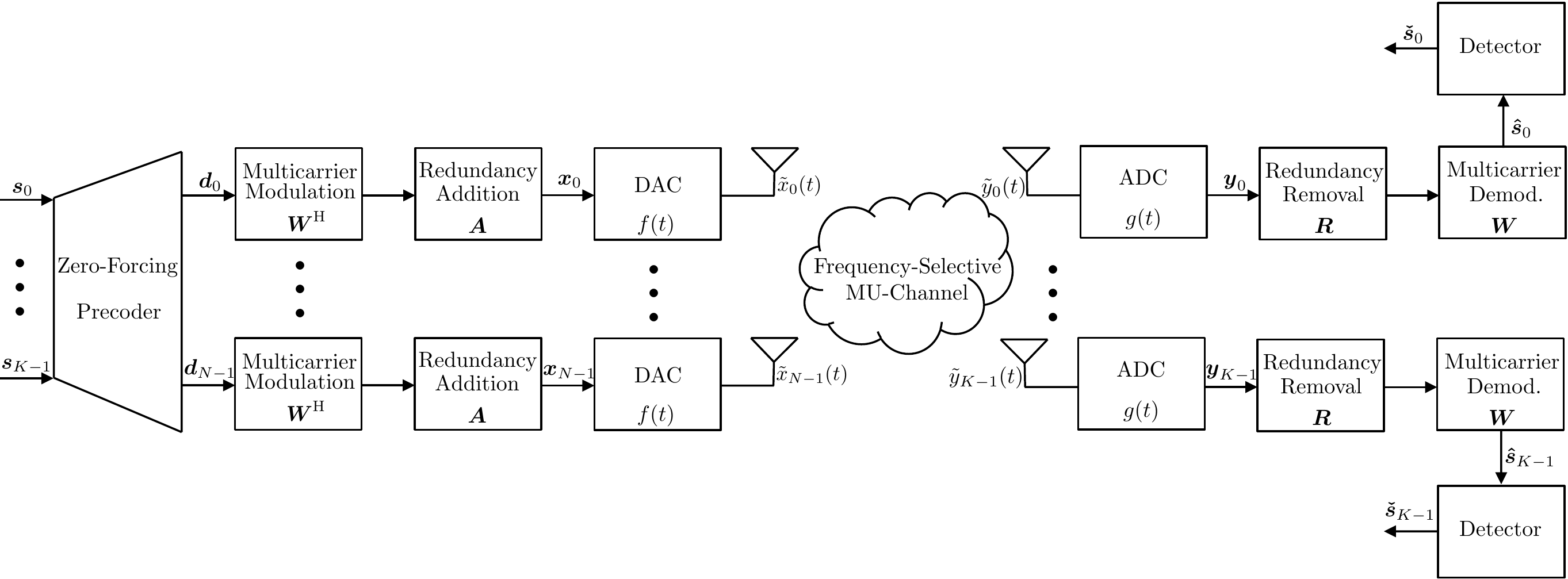}
 \caption{Multi-user MISO base-band model of a linear zero-forcing precoder.\label{fig:zf-mc}}
 \end{figure*}
 
\subsection{System Model}\label{sub:model-zf}
 
Consider the base-band MU-MISO system model of a downlink multicarrier transmission to $K$ single-antenna user terminals via $N \geq K$ antennas illustrated in Fig.~\ref{fig:zf-mc}. 


For the precoding design, we will assume perfect CSI knowledge at the transmitter side. The system works in a block-based manner, so that the data stream to be delivered to the $k^{\rm th}$ user is divided into non-overlapping blocks with $M$ symbols from a complex-valued constellation ${\cal  C}$. Let $s_{k}[m] \in {\cal C}$, with $m\in \mathscr{M} \triangleq \{0, 1, \ldots, M-1\}$,  denote the constellation symbols of one block to be transmitted to the $k^{\rm th}$ user, and $\boldsymbol{s}_k$ denote a vector gathering the $M$ symbols of the block. All vectors $\boldsymbol{s}_k$, with  $k\in \mathscr{K}\triangleq \{0, 1, \ldots, K-1\}$, are linearly processed jointly by the zero-forcing precoder to yield the precoded vectors $\boldsymbol{d}_n \in\mathbb{C}^{M\times 1}$, with  $n\in \mathscr{N}\triangleq \{0, 1, \ldots, N-1\}$. Each precoded vector $\boldsymbol{d}_n$ passes through a digital multicarrier modulator implemented via an \emph{inverse DFT} (IDFT) operation, followed by a redundancy insertion block, thus generating the vector $\boldsymbol{x}_{n}\in\mathbb{C}^{P\times 1}$, in which $P \triangleq M+R$, with $R$ denoting the number of redundant elements added as guard interval in the form of \emph{zero padding} (ZP). 

The continuous-time signal $\tilde{x}_n(t)$ that feeds the RF chain of the $n^{\rm th}$ antenna element is the output of a \emph{digital-to-analog converter} (DAC) with sampling time $T_{\rm s}$; thus, by defining the index set $\mathscr{P} \triangleq \{0, 1, \ldots, P-1\}$, one has
\begin{align}\label{eq:tx-sig}
\tilde{x}_n(t) &\triangleq \sum_{p\in\mathscr{P}} x_{n}[p]\cdot f(t-p T_{\rm s}) 
=  \left(x_{n}\ast f\right)(t),
\end{align}
where $f(t)$ denotes the transmitting pulse (e.g., a \emph{square-root raised cosine}---SRRC) and
\begin{align}\label{eq:tx-sig-clean}
x_{n}(t) \triangleq \sum_{p\in\mathscr{P}}x_{n}[p]\cdot\delta(t-p T_{\rm s})\,.
\end{align}  


Let us assume a frequency-selective channel whose coherence time is longer than the duration for transmitting one block of signals, and let $\tilde{h}_{k,n}(t)$ denote the impulse response of the base-band physical link between the $n^{\rm th}$ transmitting antenna and the $k^{\rm th}$ user terminal. The base-band signal received by the $k^{\rm th}$ user is
\begin{align}
\tilde{y}_k(t) \triangleq \sum_{n\in\mathscr{N}}\left(\tilde{h}_{k,n}\ast\tilde{x}_n\right)(t) + \tilde{v}_k(t)\,,
\end{align}
 in which $\tilde{v}_k(t)$ is an additive noise signal. 

The signal at the output of the receiving filter $g(t)$ is
\begin{align}\label{eq:y_kt}
y_k(t) \triangleq \sum_{n\in\mathscr{N}}\sum_{p\in\mathscr{P}} x_{n}[p]\cdot h_{k,n}(t-p T_{\rm s})+v_k(t)\,,
\end{align}
 in which $v_k(t) \triangleq \left(g\ast \tilde{v}_k\right)(t)$ is the equivalent noise and
 \begin{align}\label{eq:hkn}
 h_{k,n}(t) \triangleq (\tilde{h}_{k,n}\ast g\ast  f)(t)
 \end{align} 
 is the equivalent base-band channel model. After sampling the signal $y_k(t)$, the resulting samples are collected in the vector $\boldsymbol{y}_k\in\mathbb{C}^{P\times 1}$ and then processed via a redundancy removal step, followed by a multicarrier demodulation implemented via a DFT operation, before the actual symbol detection.

 \subsection{Linear ZF Precoder}
 
 When dealing with standard linear ZF precoders, it is implicitly assumed that $R$ is sufficiently large so that  
the redundancy addition at the transmitter and removal at the receiver are able to eliminate IBI as well as to induce a circulant-channel structure for each equivalent link between transmitting antennas and user terminal. 

At the receiver side, after the redundancy removal the resulting signal is multiplied by the $M$-dimensional unitary DFT matrix $\boldsymbol{W}$, which combined with the IDFT in the transmitter, is able to diagonalize the corresponding equivalent channel matrix~\cite{Diniz2012}. Hence, one can write the estimated symbols of the $k^{\rm th}$ user as
\begin{align}
\boldsymbol{\hat{s}}_k 
&= \sum_{n\in\mathscr{N}}\boldsymbol{\Lambda}_{k,n}\boldsymbol{d}_n + \boldsymbol{z}_k \,,
\end{align}
where $\boldsymbol{\Lambda}_{k,n}$ is an $M\times M$ diagonal matrix containing the frequency response of the equivalent channel between the $n^{\rm th}$ antenna element and the $k^{\rm th}$ user terminal~\cite{Diniz2012}, and $\boldsymbol{z}_k$ is the equivalent additive noise. 

Thus, one has the following complete model:
\begin{align}\label{eq:shat_k}
\underbrace{\begin{bmatrix}\boldsymbol{\hat{s}}_0 \\ \vdots \\ \boldsymbol{\hat{s}}_{K-1}\end{bmatrix}}_{\triangleq \boldsymbol{\hat{s}}\in\mathbb{C}^{MK\times 1}} 
& =
\underbrace{\begin{bmatrix}
\boldsymbol{\Lambda}_{0,0} &\hspace*{-0.2cm} \cdots\hspace*{-0.2cm} & \boldsymbol{\Lambda}_{0,N-1}\\
\vdots &\hspace*{-0.2cm} \vdots\hspace*{-0.2cm} &  \vdots \\
\boldsymbol{\Lambda}_{K-1,0} & \hspace*{-0.2cm}\cdots\hspace*{-0.2cm} & \boldsymbol{\Lambda}_{K-1,N-1}
\end{bmatrix}}_{\triangleq \boldsymbol{H}\in\mathbb{C}^{MK\times MN}}\hspace*{-0.1cm} 
\underbrace{\begin{bmatrix}\boldsymbol{d}_0 \\ \vdots \\ \boldsymbol{d}_{N-1}\end{bmatrix}}_{\triangleq \boldsymbol{d}\in\mathbb{C}^{MN\times 1}}\hspace*{-0.1cm} +\hspace*{-0.1cm}  
\underbrace{\begin{bmatrix}\boldsymbol{z}_0 \\ \vdots \\ \boldsymbol{z}_{K-1}\end{bmatrix}}_{\triangleq \boldsymbol{z}\in\mathbb{C}^{MK\times 1}}
\end{align}
or, simply, 
\begin{align}\label{eq:final_model}
\boldsymbol{\hat{s}} &= \boldsymbol{H}\boldsymbol{d} + \boldsymbol{z}\,.
\end{align}

The ZF precoded signals are therefore obtained as~\cite{Nossek2005}
\begin{align}
\boldsymbol{d} = \boldsymbol{H}^{\rm H}\left(\boldsymbol{H}\boldsymbol{H}^{\rm H}\right)^{-1}\boldsymbol{s}\,,
\end{align}
wherein $\boldsymbol{s} \triangleq \left[\,\boldsymbol{s}_0^\intercal\;\cdots\; \boldsymbol{s}_{K-1}^\intercal\,\right]^\intercal$. Note that, due to the structure of $\boldsymbol{H}$, the signals $s_{k}[m]$ modulating the $m^{\rm th}$ subcarrier  are spatially combined for all users $k\in \mathscr{K}$ without exchanging data among subcarriers (spatial-domain precoding).  In other words, this multicarrier model is equivalent to $M$ parallel single-carrier precoded transmissions. Indeed, the $m^{\rm th}$ entry of vector $\boldsymbol{\hat{s}}_k$ in~\eqref{eq:shat_k} can be written as
\begin{align}
    \hat{s}_{k}[m] &= \left(\boldsymbol{\lambda}^{(m)}_k\right)^\intercal\boldsymbol{d}^{(m)} + z_{k}[m], \quad \forall m \in \mathscr{M},
\end{align}
where $\boldsymbol{\lambda}^{(m)}_k \in \mathbb{C}^{N\times 1}$ collects the $m^{\rm th}$ diagonal entry of all the $N$ matrices $\boldsymbol{\Lambda}_{k,n}$, with $n \in \mathscr{N}$, whereas $\boldsymbol{d}^{(m)} \in \mathbb{C}^{N\times 1}$ collects the $m^{\rm th}$ entry (corresponding to the $m^{\rm th}$ subcarrier) of all the $N$ precoded vectors $\boldsymbol{d}_{n}$, with $n \in \mathscr{N}$. Now, considering all the $K$ users, one gets
\begin{align}
\underbrace{\begin{bmatrix}\hat{s}_{0}[m] \\ \vdots \\ \hat{s}_{K-1}[m]\end{bmatrix}}_{\triangleq \boldsymbol{\hat{s}}^{(m)}\in\mathbb{C}^{K\times 1}} 
& = 
\underbrace{\begin{bmatrix}
\left(\boldsymbol{\lambda}^{(m)}_0\right)^\intercal\\
\vdots\\
\left(\boldsymbol{\lambda}^{(m)}_{K-1}\right)^\intercal
\end{bmatrix}}_{\triangleq \boldsymbol{H}^{(m)}\in\mathbb{C}^{K\times N}}\boldsymbol{d}^{(m)} + 
\underbrace{\begin{bmatrix}z_{0}[m] \\ \vdots \\ z_{K-1}[m]\end{bmatrix}}_{\triangleq \boldsymbol{z}^{(m)}\in\mathbb{C}^{K\times 1}}, \quad \forall m \in \mathscr{M},
\end{align}
or, simply, 
\begin{align}\label{eq:final_model_subcarrier}
\boldsymbol{\hat{s}}^{(m)} &= \boldsymbol{H}^{(m)}\boldsymbol{d}^{(m)} + \boldsymbol{z}^{(m)}, \quad \forall m \in \mathscr{M}\,.
\end{align}
Therefore, the (per-subcarrier) ZF precoded signals can alternatively be written as
\begin{align}
\boldsymbol{d}^{(m)} = \left(\boldsymbol{H}^{(m)}\right)^{\rm H}\left[\boldsymbol{H}^{(m)}\left(\boldsymbol{H}^{(m)}\right)^{\rm H}\right]^{-1}\boldsymbol{s}^{(m)}, \quad \forall m \in \mathscr{M}\,,
\end{align}
wherein $\boldsymbol{s}^{(m)} \triangleq \left[\,s_0[m]\;\cdots\; s_{K-1}[m]\,\right]^\intercal$.

\subsection{Remarks}\label{sub:rem-ZF}

Usually, $(f\ast g)(t)$ is a $T_1$-Nyquist filter with roll-off factor $\rho \in (0,1]$ and period $T_1 > 0$, so that no additional ISI is induced as long as: (i) $T_{\rm s} = T_1$; (ii) the received signal employs the same sampling time $T_{\rm s}$; and (iii) there is no time offset. Indeed, $(f\ast g)(pT_1) \propto \delta[p-\delta_0]$, where $\delta_0$ is a discrete-time delay; this means the transmitting/receiving filters spread the signal in time without inducing interference among the samples spaced apart at multiples of $T_1$. 
However, when attempting to increase throughput by increasing the baud-rate making $T_{\rm s} < T_1$, one may end up with an equivalent channel for which the number of redundant elements $R$ is not sufficient to guarantee complete IBI elimination. In fact, the residual interference might be significant enough to impair the overall system performance in terms of bit-error rate or achievable data rate.  

The next section addresses the case in which the transmission of signals are frequency-packed by using fractional Fourier transform and accelerated by using a sampling time $T_{\rm s} < T_1$.

%

\section{Frequency-packed Faster-than-Nyquist Signaling}\label{sec:ftn}

Let us start by analyzing the spectrum of the transmitted signals for the system model in Section~\ref{sub:model-zf}. The Fourier transform of the base-band transmitted signal in~\eqref{eq:tx-sig} is
\begin{align}
\tilde{X}_n({\rm j}\Omega) &=  X_n\left({\rm e}^{{\rm j}\Omega T_{\rm s}}\right)F({\rm j}\Omega),
\end{align}
where $X_n\left({\rm e}^{{\rm j}\omega}\right)$ is the \emph{discrete-time Fourier transform} (DTFT) of the sequence $x_{n}[m]$, with $m\in\mathscr{M}$. When expressing the corresponding continuous-time Fourier transform, one has that $X_n\left({\rm e}^{{\rm j}\Omega T_{\rm s}}\right)$ is $\frac{2\pi}{T_{\rm s}}$-periodic, so that one can state that the entire information about the signal is, for instance, within the low-pass band $\left[0, \frac{2\pi}{T_{\rm s}}\right)$~rad/s. Being more specific, one has
\begin{align}
X_n\left({\rm e}^{{\rm j}\omega}\right) &= \sum_{m'\in\mathscr{M}}d_n[m']{\rm DTFT}\left\{\frac{1}{\sqrt{M}}{\rm e}^{{\rm j}\frac{2\pi}{M}mm'}\right\}\,,
\end{align}
and by defining the window $w[m] = \frac{1}{\sqrt{M}}$, for all $m\in\mathscr{M}$, and $w[m] = 0$ otherwise, then one can write
\begin{align}
{\rm DTFT}\left\{w[m]{\rm e}^{{\rm j}\frac{2\pi}{M}mm'}\right\} &= W\left({\rm e}^{{\rm j}\left(\omega-\frac{2\pi}{M}m'\right)}\right),\quad \forall m' \in \mathscr{M}\,,
\end{align}
where 
\begin{align}
W\left({\rm e}^{{\rm j}\omega}\right) \triangleq \frac{{\rm e}^{-{\rm j}\omega\left(\frac{M-1}{2}\right)}}{\sqrt{M}}\frac{\sin\left(\frac{\omega M}{2}\right)}{\sin\left(\frac{\omega}{2}\right)}.
\end{align}

Although the support of the $2\pi$-periodic function $W\left({\rm e}^{{\rm j}\omega}\right)$ spans, for instance, the whole interval $[0, 2\pi)$~rad/sample, let us assume for the sake of discussion that the main lobe of $W\left({\rm e}^{{\rm j}\omega}\right)$ essentially defines its bandwidth, which is therefore $\frac{2\pi}{M}$. Thus, the bandwidth of $X_n\left({\rm e}^{{\rm j}\omega}\right)$ in terms of normalized frequency in the range $[0, 2\pi)$~rad/sample is
\begin{align}
\frac{2\pi}{M}(M-1) + \frac{2\pi}{M} = 2\pi\,,
\end{align}
which shows that, no matter the number of subcarriers $M$, the MC-modulated signal is full-band and, therefore, the continuous-time signal before the pulse shaping, $x_n(t)$, occupies the entire band $\left[0, \frac{2\pi}{T_{\rm s}}\right)$~rad/s.  

\subsection{Frequency-packed Transmissions}

One can attempt to increase the spectrum efficiency of the MC-modulated scheme in Fig.~\ref{fig:zf-mc} by replacing the  IDFT matrix in the transmitter side with the so-called \emph{inverse fractional Fourier transform} (IFrFT) matrix $\boldsymbol{W}_{\beta}^{\rm H}$~\cite{Bailey1991}---see the proposed transmitter structure in Fig.~\ref{fig:slp-mc} on p.~\pageref{fig:slp-mc}. The entry $(m,m')\in\mathscr{M}^2$ of the IFrFT matrix is ${\rm e}^{{\rm j}\frac{2\pi \beta}{M}mm'}$, wherein $\beta \in (0, 1]$ is a parameter that controls the frequency packing. In addition, considering a post-IFrFT modulation by $w[m]{\rm e}^{-{\rm j}\pi \beta\left(\frac{M-1}{M}\right) m}$, the entry $(m,m)$ of the diagonal matrix $\boldsymbol{\Sigma}_\beta$ in Fig.~\ref{fig:slp-mc}, which shifts the frequency-domain content so that the center frequency is now at $0$~rad (instead of $\pi$~rad),\footnote{The modulation by ${\rm e}^{-{\rm j}\pi \beta\left(\frac{M-1}{M}\right) m}$ is particularly important when accelerating the transmissions via FTN, since the magnitude response of the transmitting filter, $|F({\rm j}\Omega)|$, is  symmetric around the origin.}  then  one has
\begin{align}
X_n\left({\rm e}^{{\rm j}\omega}\right) &= \sum_{m'\in\mathscr{M}}d_n[m']{\rm DTFT}\left\{w[m]{\rm e}^{-{\rm j}\pi \beta\left(\frac{M-1}{M}\right) m}{\rm e}^{{\rm j}\frac{2\pi\beta}{M}mm'}\right\}\,,
\end{align}
where 
\begin{align}
{\rm DTFT}\left\{w[m]{\rm e}^{-{\rm j}\pi \beta\left(\frac{M-1}{M}\right) m}{\rm e}^{{\rm j}\frac{2\pi\beta}{M}mm'}\right\} &= W\left({\rm e}^{{\rm j}\left(\omega-\beta\frac{2\pi}{M}(m'-(M-1)/2)\right)}\right),\quad \forall m' \in \mathscr{M}\,,
\end{align}
so that the bandwidth of $X_n\left({\rm e}^{{\rm j}\omega}\right)$ in the range $[-\pi, \pi)$~rad/sample is now
\begin{align}\label{eq:xiDef}
\beta\frac{2\pi}{M}(M-1) + \frac{2\pi}{M} = 2\pi\cdot\underbrace{\left(\frac{M-1}{M}\beta+\frac{1}{M}\right)}_{\triangleq \xi_M(\beta)} = 2\pi\xi_M(\beta)\,,
\end{align}
in which $\xi_M(\beta) < 1$ for any $\beta \in(0,1)$ as long as $M > 1$, whereas $\xi_M(1) = 1$ for any $M\geq 1$. Therefore, the continuous-time signal before the pulse shaping, $x_n(t)$, essentially occupies the band $\left[-\frac{\pi\xi_M(\beta)}{T_{\rm s}}, \frac{\pi\xi_M(\beta)}{T_{\rm s}}\right)$~rad/s. 
The transmitting and receiving filters can thus be chosen as identical pulses (i.e., $g(t) = f(t)$), such that $|F({\rm j}\Omega)|^2$ satisfies the  $T_{\beta}$-Nyquist ISI-free property~\cite{PP2010}, with 
\begin{align}
T_{\beta} = \frac{T_1}{\xi_M(\beta)}\,,
\end{align}
and with frequency support $\left[-\frac{(1+\rho)\pi\xi_M(\beta)}{T_1}, \frac{(1+\rho)\pi\xi_M(\beta)}{T_1}\right)$~rad/s.

\subsection{Faster-than-Nyquist Transmissions}

When $T_{\rm s} = \alpha T_1 < T_{\beta} = T_1\xi_M(\beta)$, the transmissions of samples are accelerated using an FTN signaling~\cite{Anderson2013}. This happens when $\alpha\xi_M(\beta)\in (0, 1)$. 
In this case, the equivalent channel $h_{k,n}^{(\beta)}(t)$ in~\eqref{eq:hkn}---now with the explicit dependency on the factor $\beta$, since $(f\ast g)(t)$ is a $T_{\beta}$-Nyquist filter---accounts for the ISI induced by the fact that the Nyquist pulse assumption is no longer valid, thus potentially impacting the parameters' choice of the system, such as the guard-interval length $R$ defined in Section~\ref{sub:model-zf}.

On the one hand, $\alpha$ should be made small to accelerate as much as possible the transmissions aiming to achieve higher data rates. On the other hand, $\alpha \leq 1$ should be made large to reduce the harmful ISI effects over the transmissions; notice also that, when there is significant ISI, bandwidth resources could also be spent in the transmission of redundant elements to cope with the underlying IBI, thus decreasing the spectral efficiency~\cite{Diniz2012}. Hence, it is not straightforward to guarantee that, by decreasing $\alpha$, one has  spectral efficiency gains for the system model in Section~\ref{sub:model-zf}; nor is clear to which extent  $\alpha$ can be decreased without significantly harming the transmission performance. These aspects will be further investigated in the next section.

\subsection{Theoretical Analysis}

Assume that $T_{\rm s} = \alpha T_1$ and that the discrete-time equivalent channel model of the MU-MISO system in Section~\ref{sub:model-zf} is parameterized by the coefficients $h_{k,n}[0]$, $h_{k,n}[1]$, $\ldots$, $h_{k,n}[\nu_{\alpha,\beta}] \in \mathbb{C}$, for all $(k,n)\in\mathscr{K}\times\mathscr{N}$, wherein
\begin{align}
\nu_{\alpha,\beta} \triangleq \sup_{(k,n)\in\mathscr{K}\times\mathscr{N}}\left\{\nu_0 \in\mathbb{N}\,;\, |h_{k,n}[\nu_0]| = |h_{k,n}^{(\beta)}(\alpha\nu_0T_1)| > \varepsilon\right\}
\end{align} 
for a given threshold $\varepsilon \geq 0$.\footnote{Note that $\nu_{\alpha,\beta} < \infty$ for physically meaningful channel models, otherwise one would end up with $\sum_{\nu \in \mathbb{N}}|h_{k,n}[\nu]|^2 = \infty$.} In this case, $\nu_{\alpha,\beta}$ denotes the order of the MU-MISO channel for the sampling time $T_{\rm s} = \alpha T_1$ and $T_{\beta}$-Nyquist filter $(f\ast g)(t)$. 

Note that, for another sampling time $T_{\rm s}' = \alpha' T_1$, with $\alpha' \in (0,1]$, one has $\nu_{\alpha,\beta} \approx \left(\frac{\alpha'}{\alpha}\right)\cdot\nu_{\alpha',\beta}$. As the number of redundant elements $R_{\alpha,\beta}$---now with the explicit dependency on the factors $\alpha$ and $\beta$---is usually proportional to $\nu_{\alpha,\beta}$~\cite{Lin2002,Diniz2012}, let us assume that $R_{\alpha,\beta} = \left\lfloor\frac{\alpha'}{\alpha}\cdot R_{\alpha',\beta}\right\rfloor$, with $\left\lfloor \cdot \right\rfloor $ standing for the floor function. Intuitively, the tendency is that when transmissions are accelerated, the channel order increases along with the corresponding guard-interval length.\footnote{In other words, in order to save bandwidth for useful data, one parsimoniously increases the amount of redundant elements---hence the use of the floor function to obtain an integer value---when accelerating the transmissions using $0 < \alpha < \alpha' \leq 1$.} Also, when more symbols are packed in the frequency domain (i.e., when $\beta$ decreases), the effect of the underlying $T_{\beta}$-Nyquist filter $(f\ast g)(t)$ upon the physical channel $\tilde{h}_{k,n}(t)$ in~\eqref{eq:hkn} is to focus on a narrower channel band, thus potentially yielding discrete-time channel models with lower orders, requiring less redundant elements in the transmission. 


Furthermore, consider the following definition corresponding to the achievable data rate normalized by the bandwidth.
\begin{definition}\label{defn:SE0}
The \emph{error-free spectral efficiency} is
\begin{align}
{\rm SE}_0(\alpha,\beta) \triangleq \frac{M}{M+R_{\alpha,\beta}}\cdot \frac{1}{\alpha\xi_M(\beta)}\cdot \frac{b\cdot r_{\rm c}}{2(1+\rho)}\quad[\rm bit/s/Hz],
\end{align}
in which  $\xi_M(\beta)$ is defined in~\eqref{eq:xiDef}, $b$ is the number of bits per constellation symbol, $r_{\rm c}$ is the channel coding rate, and $\rho \in (0, 1]$ is the roll-off factor.
\end{definition}

With this definition, one has the following result.
\begin{proposition}\label{prop:seGainsAlpha}
Given $\beta \in (0,1]$, ${\rm SE}_0(\alpha,\beta)$ is a decreasing function of $\alpha \in (0, 1]$.
\begin{proof}
The proof is given in Appendix~\ref{proof:prop:seGainsAlpha}. \end{proof}
\end{proposition}

\begin{remark}
In fact, from the proof of Proposition~\ref{prop:seGainsAlpha}, one can be more precise and state that the relative gain in the error-free spectral efficiency that one can get from decreasing $\alpha'$ to $\alpha$ is
\begin{align}
\frac{{\rm SE}_0(\alpha,\beta)-{\rm SE}_0(\alpha',\beta)}{{\rm SE}_0(\alpha',\beta)} \geq \left[1-\left(\frac{\alpha}{\alpha'}\right)\right]\frac{M}{M+R_{\alpha',\beta}}\,.
\end{align}
\end{remark}

\begin{proposition}\label{prop:seGainsBeta}
Given $\alpha \in (0,1]$, ${\rm SE}_0(\alpha,\beta)$ is a decreasing function of $\beta \in (0, 1]$.
\begin{proof}
This follows straightforwardly from $R_{\alpha,\beta}$ being a monotonic non-increasing function of $\beta$ and $\xi_M(\beta)$ being a monotonic decreasing function of $\beta$. \end{proof}
\end{proposition}

\begin{remark}
Note that the results in Propositions~\ref{prop:seGainsAlpha} and~\ref{prop:seGainsBeta} still hold when $R_{\alpha,\beta}$ is constant (i.e., independent of $\alpha$ and $\beta$). This case is also common in some practical systems that define a fixed  guard-interval length based on CSI statistics, instead of optimizing it for the instantaneous CSI.  
\end{remark}


These results suggest that one should decrease $T_{\rm s}$ as much as possible or increase $T_{\beta}$ as much as possible; yet, when doing so, the channel taps $h_{k,n}[p]$ also change (even if the physical link is kept the same). This eventually means one cannot discard the possibility of having an equivalent channel more favorable to the transmission (performance-wise, so to speak) when using, for instance,  a larger $T_{\rm s}$, which eventually might impact the \emph{effective} spectral efficiency---the one considering transmission errors. 
In fact, one can already infer that the degree of acceleration is a function of the roll-off factor $\rho \in(0,1]$, considering  that $|F({\rm j}\Omega)|^2$ is a $T_{\beta}$-Nyquist pulse. Indeed, $F({\rm j}\Omega)$ preserves the signal spectrum $X_n\left({\rm e}^{{\rm j}\Omega T_{\rm s}}\right)$ unchanged within the band $\left[-\frac{(1-\rho)\pi}{T_{\beta}}, \frac{(1-\rho)\pi}{T_{\beta}}\right)$~rad/s when $T_{\rm s} = T_1$, which means that virtually the transmitting pulse does not introduce  frequency selectivity in the process for small $\rho$, as long as the signal is frequency packed by a factor $\beta$. 
Now, if an acceleration factor $\alpha \in (0,1)$ is employed, then the information in $X_n\left({\rm e}^{{\rm j}\alpha\Omega T}\right)$ is within the band $\left[-\frac{\xi_M(\beta)}{\alpha}\frac{\pi}{T_1}, \frac{\xi_M(\beta)}{\alpha}\frac{\pi}{T_1}\right)$~rad/s, which means that there is an expansion of its analog frequency range. This means that part of the frequency content will be attenuated by the low-pass filter $F({\rm j}\Omega)$. The smaller the value of $\alpha$, the greater the induced frequency selectivity. Similarly, the smaller the value of $\beta$, the greater the induced  ICI due to the loss of orthogonality among subcarriers. If $\alpha$ is too small, parts of the signal spectrum $X_n\left({\rm e}^{{\rm j}\alpha\Omega T}\right)$ may be lost due to the low-pass filtering, since the output of the DAC is ideally zeroed for $|\Omega| \geq \frac{(1+\rho)\pi\xi_M(\beta)}{T_1}$. This confirms that one might accelerate the transmissions and still be able to reconstruct the signal, as long as it is above a given lower-bound, $\alpha_{\min}\in (0,1]$, for $\alpha$ to guarantee {information losslessness}~\cite{Martins2020} in the following sense.

\begin{definition}\label{defn:infoLossless}
The pair $(f(t), T_{\rm s})$ yields an \emph{information-losslessness transmission} if, given $\tilde{x}_n(t)$ in~\eqref{eq:tx-sig}, there exists a receiving filter $g(t)$ such that $\hat{x}_n[p] \triangleq \hat{x}_n(pT_{\rm s}) = x_n[p]$, $\forall p \in \mathscr{P}$, with $\hat{x}_n(t) \triangleq \left(\tilde{x}_n\ast g\right)(t)$.
\end{definition}

The following lemma characterizes the transmitting filters that enable information-losslessness transmissions.
\begin{lemma}\label{lem:minAlpha}
The information-losslessness condition in Definition~\ref{defn:infoLossless} is met if and only if
\begin{align}
\sum_{i \in \mathbb{Z}}\left\vert F\left({\rm j}\frac{(\omega + 2\pi i)}{T_{\rm s}}\right)\right\vert^2  > 0, \quad\forall \omega \in \mathbb{R}\,.
\end{align}
\begin{proof}
See Lemmas 5.1 and 5.2 in~\cite{PP2010}.
\end{proof}
\end{lemma}

Assuming a square-root $T_\beta$-Nyquist transmitting filter with roll-off factor $\rho \in (0, 1]$, one has the following result.
\begin{proposition}\label{prop:minAlpha}
Given $\beta \in(0,1]$ and $M\geq 1$, one has $\alpha_{\min} = \frac{1}{(1+\rho)\xi_M(\beta)}$.
\begin{proof}
The proof is given in Appendix~\ref{proof:prop:minAlpha}. 
\end{proof}
\end{proposition}

The next section describes how the redundancy addition/removal can deal with IBI effects.

\section{Multicarrier Symbol-level Precoding}\label{sec:slp-mc}

As explained in Section~\ref{sub:rem-ZF}, the guard interval that is usually considered for block-based transmissions might not be long enough to deal with the IBI effects stemming from accelerated signaling. As a matter of fact, even the widespread assumption that a given received data block suffers the interference from up two adjacent transmitted blocks is usually unrealistic for delay-constrained applications---i.e., those employing small to moderate block lengths $P_{\alpha,\beta} = M + R_{\alpha,\beta}$, as compared to the channel order $\nu_{\alpha,\beta}$---in which typical transmitting/receiving filters, like SRRC, are used. 
For this reason, we shall first propose an appropriate model for sequential transmissions, and then move on to the proposed system architecture for multicarrier SLP. 
 
\subsection{Sequential Transmissions} 

In sequential transmissions, the $\ell^{\rm th}$ received block after sampling, synchronization, and buffering can be written as 
\begin{align}\label{eq:y-ISI-IBI}
\boldsymbol{y}_k[\ell] \!=\!& \sum_{n\in\mathscr{N}}\!\left(\boldsymbol{H}_{{\rm ISI}_{k,n}}\!\boldsymbol{x}_n[\ell]+\sum_{b\in\mathscr{B}^{\rm (f)}}\boldsymbol{H}_{{\rm IBI}_{k,n}}^{\rm (f)}\![b]\boldsymbol{x}_n[\ell-b]+\sum_{b\in\mathscr{B}^{\rm (b)}}\boldsymbol{H}_{{\rm IBI}_{k,n}}^{\rm (b)}\![b]\boldsymbol{x}_n[\ell+b]\!\right) \!+\! \boldsymbol{v}_k[\ell]\,,
\end{align}
where $\boldsymbol{x}_n[\ell] = \left[\,x_n[0+(\ell-1)P]\;x_n[1+(\ell-1)P]\,\cdots\,x_n[P-1+(\ell-1)P]\,\right]^\intercal$, the index sets $\mathscr{B}^{\rm (f)}$ and $\mathscr{B}^{\rm (b)}$ contain positive integer numbers, whereas the ISI and forward/backward IBI\footnote{Taking $\ell$ as the time index of the current data block, the forward IBI corresponds to the interference coming from previous data blocks toward the current block (``$\ell - b \rightsquigarrow \ell$''), whereas the backward IBI corresponds to the interference that, for instance, the current block will generate toward previously transmitted blocks (``$\ell \rightsquigarrow \ell - b$'').} matrices are $P\times P$ Toeplitz matrices, in which, for a given group delay $\delta\in\mathbb{N}$ and a given pair of row-column indexes  $(p_{\rm r},p_{\rm c}) \in \mathscr{P}^2$, one has\footnote{For notation clarity's sake, the dependency on $\alpha$ and $\beta$ will be omitted in this discussion.}
\begin{subequations}
\label{eq:H-def}
\begin{align}
\label{subeq:Hisi-def}
\left[ \boldsymbol{H}_{{\rm ISI}_{k,n}} \right]_{p_{\rm r},p_{\rm c}}  &\triangleq \left\{ {\begin{array}{*{20}c}
   {h_{k,n}[p_{\rm r} - p_{\rm c}+ \delta] ,} & {0 \le p_{\rm r} - p_{\rm c} + \delta \le \nu ,}  \\
   {0,} & \textrm{otherwise.}  \\
\end{array}} \right.\,\\
\label{subeq:HibiB-def}
\left[ \boldsymbol{H}_{{\rm IBI}_{k,n}}^{\rm (f)}\![b] \right]_{p_{\rm r},p_{\rm c}}  &\triangleq \left\{ {\begin{array}{*{20}c}
   {h_{k,n}[bP + p_{\rm r} - p_{\rm c}+ \delta] ,} & {0 \le bP  + p_{\rm r} - p_{\rm c} + \delta \le \nu ,}  \\
   {0,} & \textrm{otherwise.}  \\
\end{array}} \right.\,\\
\label{subeq:HibiF-def}
\left[ \boldsymbol{H}_{{\rm IBI}_{k,n}}^{\rm (b)}\![b] \right]_{p_{\rm r},p_{\rm c}}  &\triangleq \left\{ {\begin{array}{*{20}c}
   {h_{k,n}[-bP + p_{\rm r} - p_{\rm c}+ \delta] ,} & {0 \le -bP  + p_{\rm r} - p_{\rm c} + \delta \le \nu ,}  \\
   {0,} & \textrm{otherwise.}  \\
\end{array}} \right.
\end{align}
\end{subequations}

In this context, it is important to determine how many blocks affect the $\ell^{\rm th}$ received block and classify them as forward and backward IBI, i.e., to characterize the index sets  $\mathscr{B}^{\rm (f)}$ and $\mathscr{B}^{\rm (b)}$. The following result addresses this problem.
\begin{proposition}\label{prop:noBlks}
Let the group delay be written as 
\begin{align}\label{eq:delay}
\delta = q_{\delta}P + \rho_{\delta} < \nu\,,
\end{align} 
with $q_{\delta}\in\mathbb{N}$ and $\rho_{\delta}\in \mathscr{P}$, and consider the integer number 
\begin{equation}
B \triangleq \left\lceil\frac{\nu}{P}\right\rceil + 2,
\label{eq:M-def}
\end{equation}
with $\left\lceil \cdot \right\rceil $ standing for the ceiling function. The maximum number of blocks that may affect the reception of the $\ell^{\rm th}$ data block in the sequential (streaming-like) transmission in~\eqref{eq:y-ISI-IBI} is either 
\begin{itemize}
    \item $B$ when $\rho_{\delta} \in \{1, 2, \dots, \nu-(B-3)P-1\}$, or 
    \item $B-1$ otherwise.  
\end{itemize}
 Moreover, the index sets for the forward and backward data blocks are respectively given by
\begin{align}
\mathscr{B}^{\rm (f)} = \{1,\dots,B-2-q_{\delta}\}\quad\text{and}\quad\mathscr{B}^{\rm (b)} = \{1,\dots,q_{\delta}+1\}\,.
\end{align}
\begin{proof}
The proof is given in Appendix~\ref{proof:prop:noBlks}. 
\end{proof}
\end{proposition}

\subsection{Proposed System Model}

 \begin{figure*}[!t]
\includegraphics[width=\linewidth]{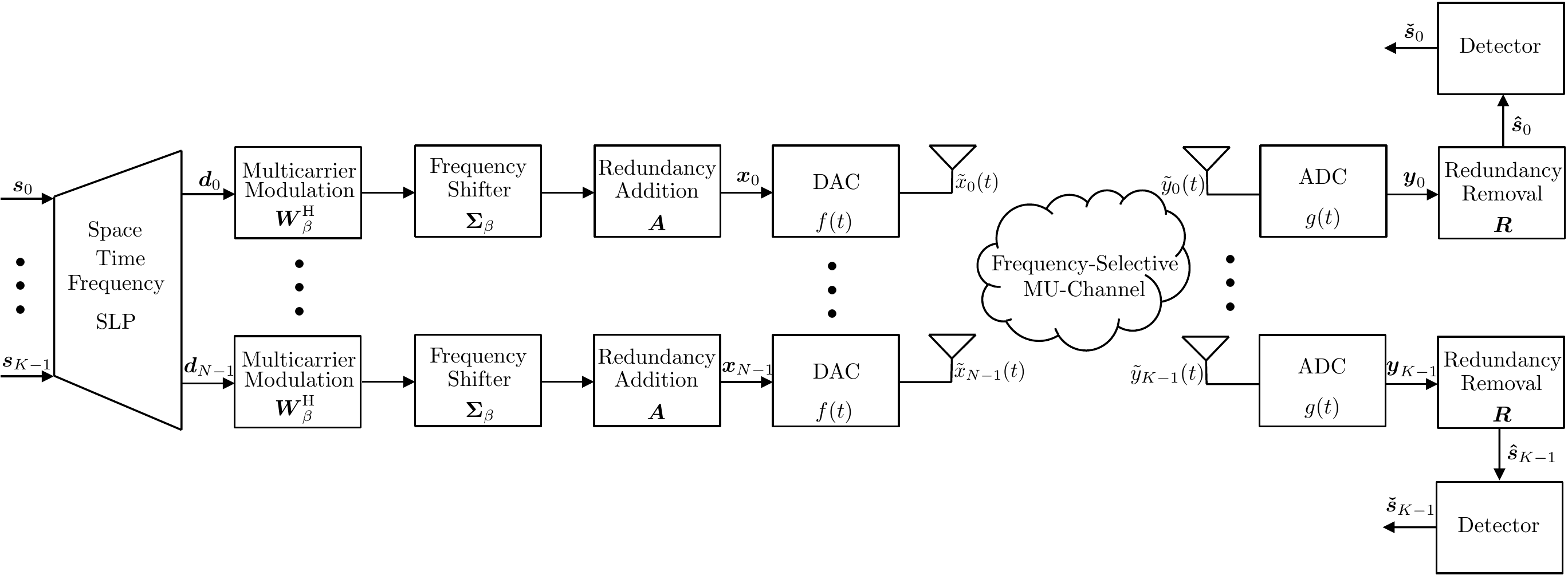}
 \caption{Proposed multi-user MISO base-band model of a space-time-frequency symbol-level precoder.\label{fig:slp-mc}}
 \end{figure*}

We propose the multi-user system architecture depicted in Fig.~\ref{fig:slp-mc}. As compared to Fig.~\ref{fig:zf-mc}, the main differences are: (i) the amount of redundancy added at the transmitter; (ii) the way the redundancy is removed at the receiver; (iii) the absence of a multicarrier digital demodulator at the receiver; (iv) the substitution of the IDFT by the IFrFT as multicarrier modulator, followed by a frequency shifter; and (v) the use of a space-time-frequency SLP.

The primary goal of symbol-level redundancy is combating IBI. However, interference plays a twofold role in symbol-level precoders, which are able to partially benefit from constructive interference---IBI included. More specifically, as 
\begin{align}
\textrm{IBI} &= \underbrace{\sum_{n\in\mathscr{N}}\sum_{b\in\mathscr{B}^{\rm (f)}}\boldsymbol{R}\boldsymbol{H}_{{\rm IBI}_{k,n}}^{\rm (f)}\![b]\boldsymbol{A}\boldsymbol{\Sigma}_{\beta}\boldsymbol{W}^{\rm H}_{\beta}\boldsymbol{d}_n[\ell-b]}_{\textrm{forward IBI}}+\underbrace{\sum_{n\in\mathscr{N}}\sum_{b\in\mathscr{B}^{\rm (b)}}\boldsymbol{R}\boldsymbol{H}_{{\rm IBI}_{k,n}}^{\rm (b)}\![b]\boldsymbol{A}\boldsymbol{\Sigma}_{\beta}\boldsymbol{W}^{\rm H}_{\beta}\boldsymbol{d}_n[\ell+b]}_{\textrm{backward IBI}}\,,
\end{align} 
while designing a block, the forward IBI from the previous block is already known and can be used as a basis for constructive interference. However, the backward IBI of the current block toward the previous one is still destructive since the previous one has been already designed.  

Thus, in order to deal with the backward IBI harmful effects, we propose the use of redundant \emph{zero-padding zero-jamming} (ZP-ZJ) transceivers~\cite{Lin2002,Martins2010a,Martins2012a,Martins2012b,Diniz2012} with an adjustable amount of redundant elements which are added and removed through the multiplication by the matrices
\begin{subequations}\label{eq:AR}
\begin{align}
\boldsymbol{A} &\triangleq \begin{bmatrix}\boldsymbol{0}_{R_{\alpha,\beta}\times M}\\ \boldsymbol{I}_M \end{bmatrix}\,,\\
\boldsymbol{R} &\triangleq \begin{bmatrix}\boldsymbol{I}_M & \boldsymbol{0}_{M \times R_{\alpha,\beta}}\end{bmatrix}\,.
\end{align}
\end{subequations}

%
Given the pair $(\boldsymbol{A},\boldsymbol{R})$ in~\eqref{eq:AR}, the following result holds.
\begin{proposition}\label{prop:minR}
The minimum guard-interval length that enables backward-IBI-free transmissions is
\begin{align}\label{eq:minR}
R_{\alpha,\beta}^{\text{\rm IBI-free}} \triangleq \left\lceil\frac{\delta}{2}\right\rceil\,.
\end{align}
\begin{proof}
The proof is given in Appendix~\ref{proof:prop:minR}. 
\end{proof}
\end{proposition}

It is worth questioning whether the amount of redundant elements inserted in the transmission is reasonably small or not. In this context, consider the following definition. 
\begin{definition}\label{defn:effTx}
The pair $(M, R_{\alpha,\beta})$ yields an \emph{efficient transmission} if $M/(M+R_{\alpha,\beta}) > 50\%$.  
\end{definition}
In plain text, Definition~\ref{defn:effTx} states that efficient transmissions are those where the useful data symbols in a block are more than the redundant symbols. 
With this definition in mind, the following result comes as a consequence of Proposition~\ref{prop:minR}. 
 \begin{corollary}\label{cor:minReff}
For backward-IBI-free efficient transmissions one must have  $\delta \in \mathscr{P}$. 
\begin{proof}
For backward-IBI-free efficient transmissions one must have $2R_{\alpha,\beta} \geq \delta$ and $M > R_{\alpha,\beta}$. Thus,  $P_{\alpha,\beta} = M + R_{\alpha,\beta} > R_{\alpha,\beta} + R_{\alpha,\beta} \geq \delta$. 
\end{proof}
\end{corollary}
\begin{remark}
From~\eqref{eq:delay} and   Corollary~\ref{cor:minReff},   backward-IBI-free efficient transmissions are possible only when $q_{\delta} = 0$, which means that there is only one backward IBI block to be eliminated through the ZP-ZJ process, since $\mathscr{B}^{\rm (b)} = \{1\}$. 
\end{remark}

We propose employing an adjustable reduced amount of redundant elements
\begin{align}\label{eq:red-set}
R_{\alpha,\beta} \in \mathscr{R}_{\alpha,\beta} \triangleq \left\{0,1,\cdots, R_{\alpha,\beta}^{\text{\rm IBI-free}}\right\},
\end{align}
which allows us to control the degree of remaining backward IBI.

Note that the pair $(\boldsymbol{A},\boldsymbol{R})$ in~\eqref{eq:AR} does not induce a circulant structure in the matrix $\boldsymbol{R}\boldsymbol{H}_{{\rm ISI}_{k,n}}\!\boldsymbol{A}$. This means that, even if a DFT matrix were to be used at the receiver side, as in Fig~\ref{fig:zf-mc}, no diagonalization of the effective channel matrix would be obtained. For this reason, we propose simplifying the receiver while letting the symbol-level precoder to deal with the resulting effective channel model. 

The reconstructed signals of all users can be written as
\begin{align}\label{eq:final_model}
\boldsymbol{\hat{s}}[\ell] &= \boldsymbol{H}_{\rm ISI}\boldsymbol{d}[\ell]+ \sum_{b\in\mathscr{B}^{\rm (f)}}\boldsymbol{H}_{\rm IBI}[b]\boldsymbol{d}[\ell-b] + \boldsymbol{z'}[\ell]\,,
\end{align}
which corresponds to the vector stacking of the signals $\boldsymbol{\hat{s}}_k[\ell]$, with $\boldsymbol{H}_{\rm ISI}\in\mathbb{C}^{KM\times NM}$ comprising the matrices $\boldsymbol{R}\boldsymbol{H}_{{\rm ISI}_{k,n}}\!\boldsymbol{A}\boldsymbol{\Sigma}_{\beta}\boldsymbol{W}^{\rm H}_{\beta}$, for $(k,n)\in\mathscr{K}\times\mathscr{N}$, that model the transfer from $\boldsymbol{\hat{d}}_n[\ell]$ to $\boldsymbol{\hat{s}}_k[\ell]$, and $\boldsymbol{H}_{\rm IBI}[b]\in\mathbb{C}^{KM\times NM}$ comprising the matrices $\boldsymbol{R}\boldsymbol{H}_{{\rm IBI}_{k,n}}^{\rm (f)}\![b]\boldsymbol{A}\boldsymbol{\Sigma}_{\beta}\boldsymbol{W}^{\rm H}_{\beta}$, for $(k,n,b)\in\mathscr{K}\times\mathscr{N}\times\mathscr{B}^{\rm (f)}$, that model the transfer from $\boldsymbol{\hat{d}}_n[\ell-b]$ (\emph{known} at instant $l$) to $\boldsymbol{\hat{s}}_k[\ell]$, whereas $\boldsymbol{z'}[\ell]\in\mathbb{C}^{KM\times 1}$ denotes the interference-plus-noise component due to the remaining backward IBI components $\boldsymbol{R}\boldsymbol{H}_{{\rm IBI}_{k,n}}^{\rm (b)}\![b]\boldsymbol{A}\boldsymbol{\Sigma}_{\beta}\boldsymbol{W}^{\rm H}_{\beta}\boldsymbol{d}_n[\ell+b]$ (\emph{unknown} at instant $l$), for $(k,n,b)\in\mathscr{K}\times\mathscr{N}\times\mathscr{B}^{\rm (b)}$, and the noise $\boldsymbol{z}_k[\ell]$.

\subsection{Non-linear Space-time-frequency Symbol-level Precoder}

We propose using a space-time-frequency SLP scheme accounting for MUI and frequency-selective-related interference---in the form of ISI, ICI, and IBI. The idea is to exploit the interference in space, time, and frequency domains in a constructive manner. Inspired by the \emph{dirty paper coding} (DPC) principle~\cite{Costa1983}, the knowledge of the forward IBI $\sum_{b\in\mathscr{B}^{\rm (f)}}\boldsymbol{H}_{\rm IBI}[b]\boldsymbol{d}[\ell-b]$ can be taken into account in the transmission, whereas the remaining backward IBI in $\boldsymbol{z'}[\ell]$ tends to degrade the system performance. 

A possible cost function is related to the minimization of total transmit power. If only one block were to be transmitted as in~\eqref{eq:tx-sig}, then the corresponding total energy (i.e., considering all transmit antennas) would be
\begin{align}\label{eq:totalPower}
E \triangleq \sum_{n\in\mathscr{N}}\int\limits_{-\infty}^{\infty}\frac{|\tilde{x}_n(t)|^2}{Z_0}{\rm d}t 
&=\frac{1}{Z_0}\sum_{n\in\mathscr{N}}\boldsymbol{x}_n^{\rm H}\boldsymbol{C}_f\boldsymbol{x}_n\,,
\end{align}
in which $Z_0 > 0$ denotes the antenna impedance, $|\tilde{x}_n(t)|^2/Z_0$ is the instantaneous power, and $\left[\boldsymbol{C}_f\right]_{p_{\rm r},p_{\rm c}} \triangleq \left(f\ast f\right)\left((p_{\rm r}-p_{\rm c})T_{\rm s}\right)$, for all $p_{\rm r}, p_{\rm c} \in {\cal P}$. As mentioned before, $f(t)$ is assumed to be a square-root $T_{\beta}$-Nyquist filter with even symmetry around the origin $t = 0$. Since $T_{\rm s} \neq T_{\beta}$ in general, then the temporal correlations of the transmitting pulse must be taken into account when modeling the transmission power, which is proportional to the total energy. 


By defining the matrix 
\begin{align}
        \boldsymbol{\Gamma} &\triangleq \boldsymbol{I}_N \otimes \left(\boldsymbol{W}_{\beta}\boldsymbol{\Sigma}_{\beta}^*\boldsymbol{A}^\intercal\boldsymbol{C}_f\boldsymbol{A}\boldsymbol{\Sigma}_{\beta}\boldsymbol{W}_{\beta}^{\rm H}\right) \in\mathbb{C}^{NM\times NM}\,,
\end{align}
then the the proposed SLP \emph{convex} optimization problem is\footnote{The problem is strictly convex, since $\boldsymbol{\Gamma}$ is a positive-definite matrix. One can use the CVX software~\cite{Boyd2006} to solve such a problem.}
\begin{align}\label{eq:opt-slp}
\begin{aligned}
 &\underset{\boldsymbol{d}[\ell]\in\mathbb{C}^{NM\times 1}}{\text{minimize}}& & \boldsymbol{d}^{\rm H}[\ell]\boldsymbol{\Gamma}\boldsymbol{d}[\ell]\\ 
& \text{subject to}
& &\boldsymbol{H}_{\rm ISI}\boldsymbol{d}[\ell] +  \sum_{b\in\mathscr{B}^{\rm (f)}}\boldsymbol{H}_{\rm IBI}[b]\boldsymbol{d}[\ell-b] \trianglerighteq  \boldsymbol{q}\odot\boldsymbol{s}[\ell]\,,
\end{aligned}
\end{align}
in which $\boldsymbol{s}[\ell]\in\mathbb{C}^{KM\times 1}$ contains the actual intended symbols for all $K$ users, $\boldsymbol{q}\in\mathbb{C}^{KM\times 1}$ collects quality-of-service-related parameters, such as target \emph{signal-to-interference-plus-noise ratio} (SINR) $\gamma_k > 0$ and noise variance $\sigma^2_k > 0$, with $k \in \mathscr{K}$,  and $\odot$ denotes point-wise product between vectors. The operator $\trianglerighteq$ can be defined in different ways depending on the specific constellation points $\boldsymbol{s}[\ell]$; in all cases, the operator $\trianglerighteq$ defines a convex constraint set (polyhedron)---please refer to~\cite{Alodeh2016,Alodeh2017b,Spano2018a,Alodeh2018,Spano2018c,Spano2018b} for further specific details omitted here.

\section{Numerical Results}\label{sec:res}

The performance of the proposed space-time-frequency symbol-level precoders is assessed via three experiments. The first one focuses on a conventional multicarrier signaling without neither frequency packing nor FTN capabilities; the main objective of this experiment is to compare the proposed SLP approach against the conventional OFDM-based linear ZF precoder for multipath-fading channels across different number of users. The second numerical experiment focuses on frequency-packed FTN multicarrier signaling over flat-fading channels across different SINRs; the main objective here is to explore the underlying trade-offs while varying the acceleration and packing factors, $\alpha$ and $\beta$ respectively, in terms of data rate, transmit power, as well as to showcase the interplay between spectral and energy efficiencies.  The third experiment focuses on frequency-packed FTN multicarrier signaling over multipath-fading channels across different values of $\alpha$ and $\beta$. The main objective here is to verify some of the theoretical predictions regarding $\alpha$ and $\beta$ in a realistic scenario. 

For all experiments, the transmitting/receiving filters are $T_\beta$-SRRC pulses,  for a fixed basic Nyquist period $T_1 = 100$~ns, and with roll-off factor $\rho = 1/4$. The additive noise is white Gaussian with variance fixed at $\sigma^2 = 1$, and the antenna impedance employed for power calculations is $Z_0 = 50~\Omega$.  

The figures of merit are: 
\begin{enumerate}
    \item[(i)] \emph{effective sum rate (or throughput)} given by ${\rm SR} \triangleq K\cdot\frac{M}{P}\cdot \frac{b}{T_{\rm s}}\cdot(1-{\rm SER})~[\rm bit/s]$, in which ${\rm SER} \triangleq \frac{1}{K}\sum_{k\in\mathscr{K}}{\rm SER}^{(k)}$ is the average \emph{symbol error rate} and $b$ is the number of bits per constellation symbol.
    \item[(ii)] \emph{time-averaged total transmit power}, $\mathbb{P}_{\rm t}$, computed as the total analog power [W] feeding the transmitting antennas averaged over the interval $P T_{\rm s}$. Here we consider the actual analog power including the cross-power terms resulting from the sequential block-based transmission. 
    \item[(iii)] \emph{effective system spectral efficiency}, ${\rm SE} \triangleq \frac{\rm SR}{\rm BW}~[\rm bit/s/Hz]$, wherein ${\rm BW} \triangleq \frac{2(1+\rho)}{T_{\beta}}$ is the bandwidth of the base-band transmitted signal. 
    \item[(iv)] \emph{overall energy inefficiency}, ${\rm EI} \triangleq \frac{\mathbb{P}_{\rm t}}{\rm SE}~[\rm J/bit]$.
\end{enumerate}

\subsection{Experiment 1}
We first study the performance of the proposed SLP approach with respect to different number of users $K \in \{4,\ldots, N\}$, for a downlink transmission using $N = 8$ antennas. In this experiment, no frequency-packed FTN signaling is considered, i.e., $(\alpha,\beta) = (1,1)$. The number of subcarriers is fixed at $M = 64$, the constellation ${\cal C}$ is 16-QAM, and the base-band multipath-fading models correspond to Rayleigh channels with exponentially decaying power profile and with resulting order varying around $\nu = 24$. As there is no time acceleration ($\alpha = 1$), $\delta = 0$ yielded well-conditioned effective channel matrices. The quality of service vector is $\boldsymbol{q} = \sqrt{\gamma\sigma^2}\boldsymbol{1}_{KM\times 1}$, where the target SINR $\gamma$ corresponds to 15~dB. The guard-interval length is chosen as: (i) $R = 0$, called `No red. SLP' for the proposed space-time-frequency SLP,\footnote{Since $\delta = 0$, one has $\mathscr{R}_{\alpha,\beta} = \{0\}$ in~\eqref{eq:red-set}.} and (ii) $R = \nu$, called `Full red. ZF', for the baseline ZF precoder using full redundancy, which is able to both eliminate IBI and induce a circulant effective channel matrix.

\begin{figure*}[!t]
\centering
\subfigure[\footnotesize Effective sum rate.]{%
\label{fig:Exp1-rate-users}%
\includegraphics[width=.33\linewidth]{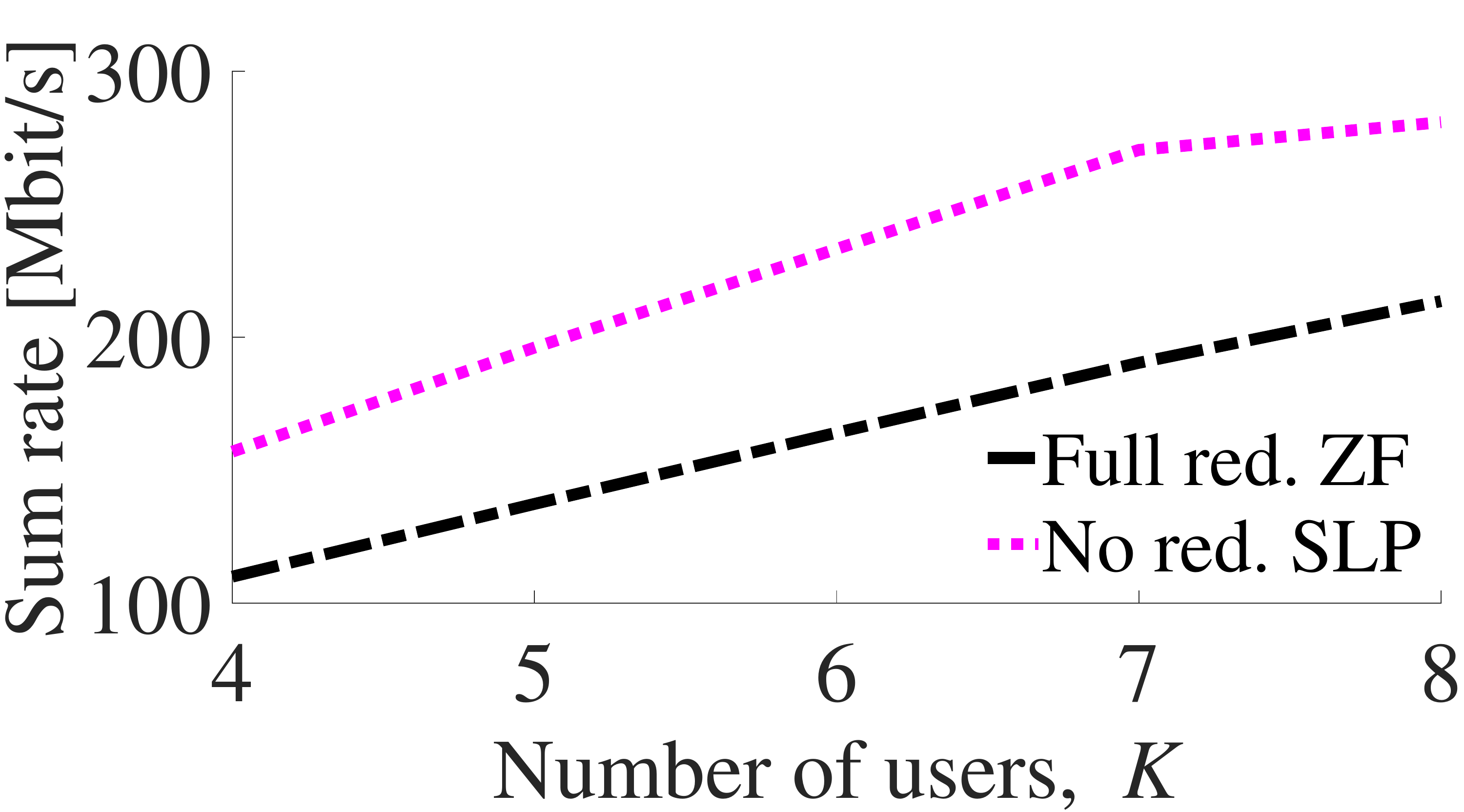}
}%
\subfigure[\footnotesize Time-averaged total transmit power.]{%
\label{fig:Exp1-power-users}%
\includegraphics[width=.33\linewidth]{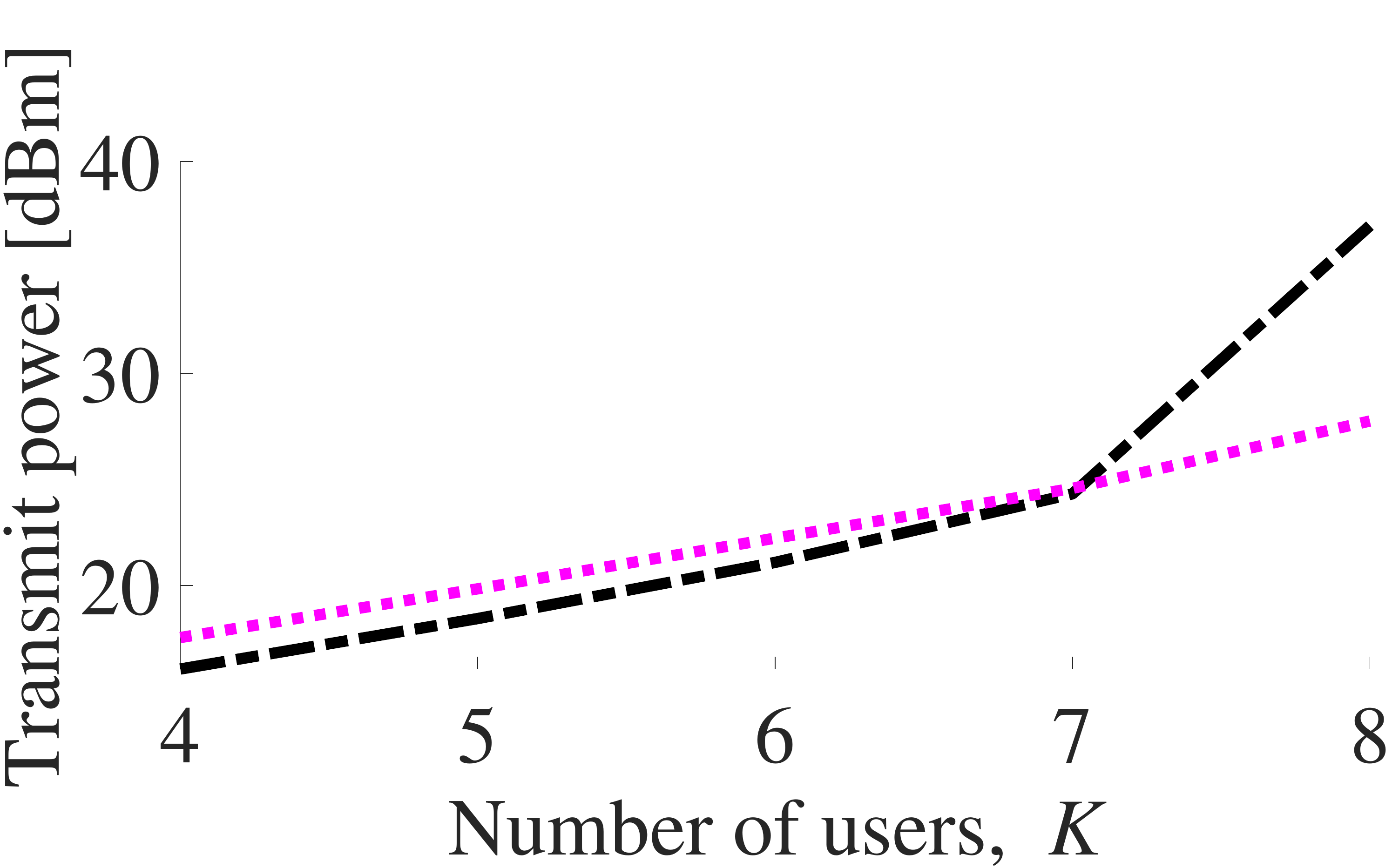}
}%
\subfigure[\footnotesize Spectral efficiency \emph{vs.} energy inefficiency for $K \in \{4,\ldots, 8\}$.]{%
\label{fig:Exp1-SEvsEI-users}%
\includegraphics[width=.33\linewidth]{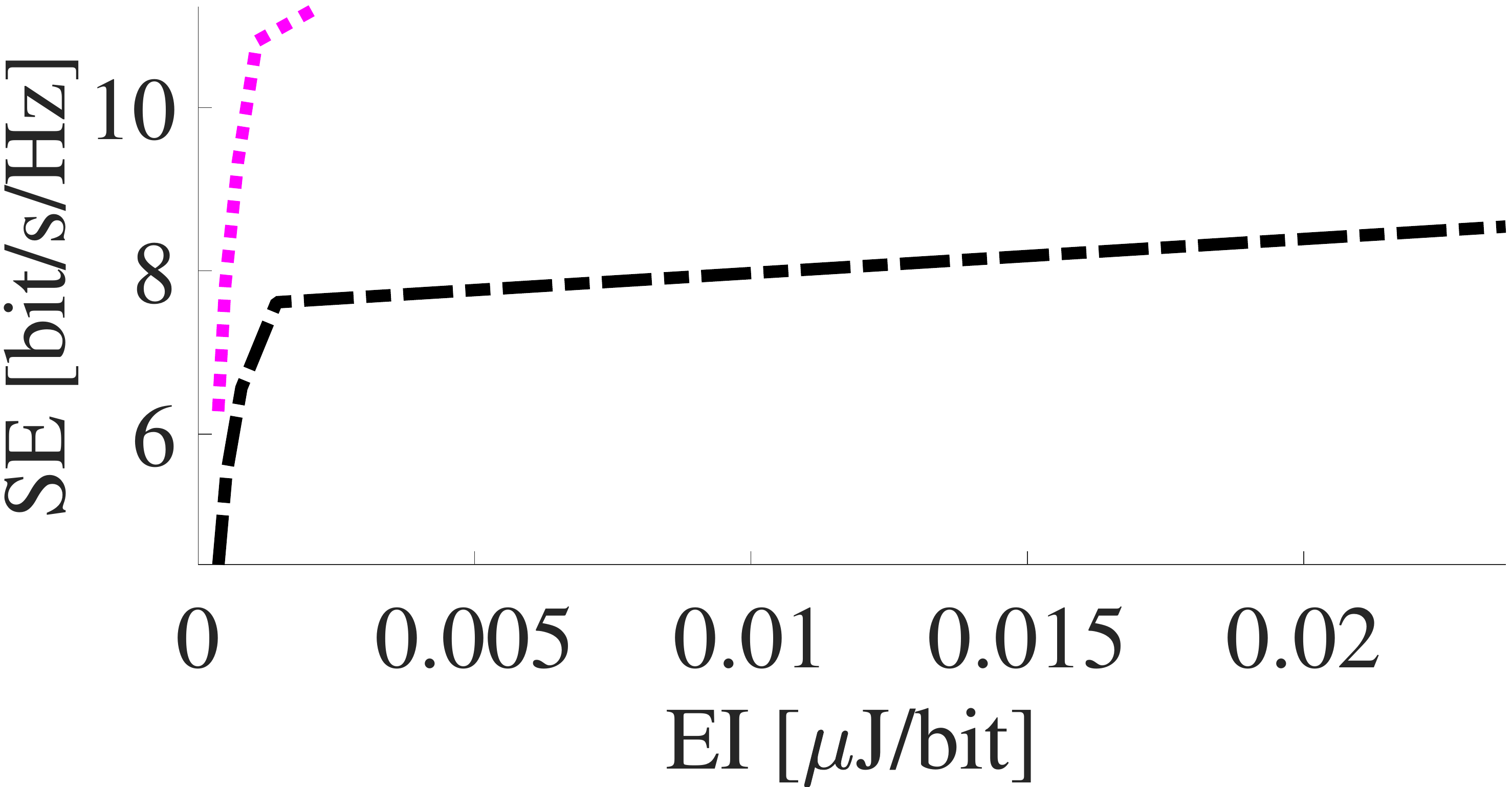}
}%
\caption{(\emph{Experiment 1}) Performance comparison between the ZF precoder with full redundancy and the proposed space-time-frequency SLP with no redundancy, as function of $K$ and for $(\alpha, \beta) = (1,1)$. The legend in (a) applies to all figures.}\label{fig:exp1}
\end{figure*}

Fig.~\ref{fig:exp1} depicts the results. Fig.~\ref{fig:Exp1-rate-users} lets clear that the proposed SLP scheme outperforms the ZF precoder in terms of effective sum rate. Note that the IBI had no harmful effect whatsoever on the  non-redundant precoder. The scenario with a higher number of users tends to be more challenging, but the SLP is able to gain more with the interference exploitation. For $K = 8$, the throughput gap drops a bit at the cost of a much higher transmit power for the ZF precoder. Indeed,  Fig.~\ref{fig:Exp1-power-users} shows that, although the ZF precoder requires slightly less transmit power than the proposed SLP for $K \leq 7$, it requires much more power in a fully-loaded system ($K = N = 8$).  Fig.~\ref{fig:Exp1-SEvsEI-users} shows the superior performance of the proposed SLP when both spectral and energy efficiencies are taken into account for all considered $K \in \{4, \ldots, N\}$.

\subsection{Experiment 2}



We now study the SLP performance for frequency-packed FTN transmissions over flat-fading channels. We consider delay-constrained applications with a short multicarrier-symbol duration (i.e., with a small number of subcarriers, $M = 16$) for both QPSK and 8PSK constellations, in a fully-loaded system with $N = K = 4$. Here we address the following questions: Is it really necessary to use frequency-packed FTN? Why not simply increasing the constellation size (e.g., from QPSK to 8PSK)? What are the trade-offs when we change the parameters $\alpha$ and $\beta$?
For the sake of clarity, we consider only the `No red. SLP' for different values of $\alpha$ and $\beta$. More specifically, $(\alpha,\beta)\in\{(0.90,0.88), (0.80,1), (1,0.79), (1,1)\}$.\footnote{Note that, when $(\alpha,\beta)\in\{(0.90,0.88), (0.80,1), (1,0.79)\}$, one has $\alpha\cdot\xi_M(\beta) \approx \frac{1}{1+\rho}$; see also Proposition~\ref{prop:minAlpha}.} 

\begin{figure*}[!t]
\centering
\subfigure[\footnotesize Spectral efficiency \emph{vs.} energy inefficiency for target SINR in $\left(-2, 12\right)$~dB.]{%
\label{fig:Exp3-SEvsEI-gamma}%
\includegraphics[width=.45\linewidth]{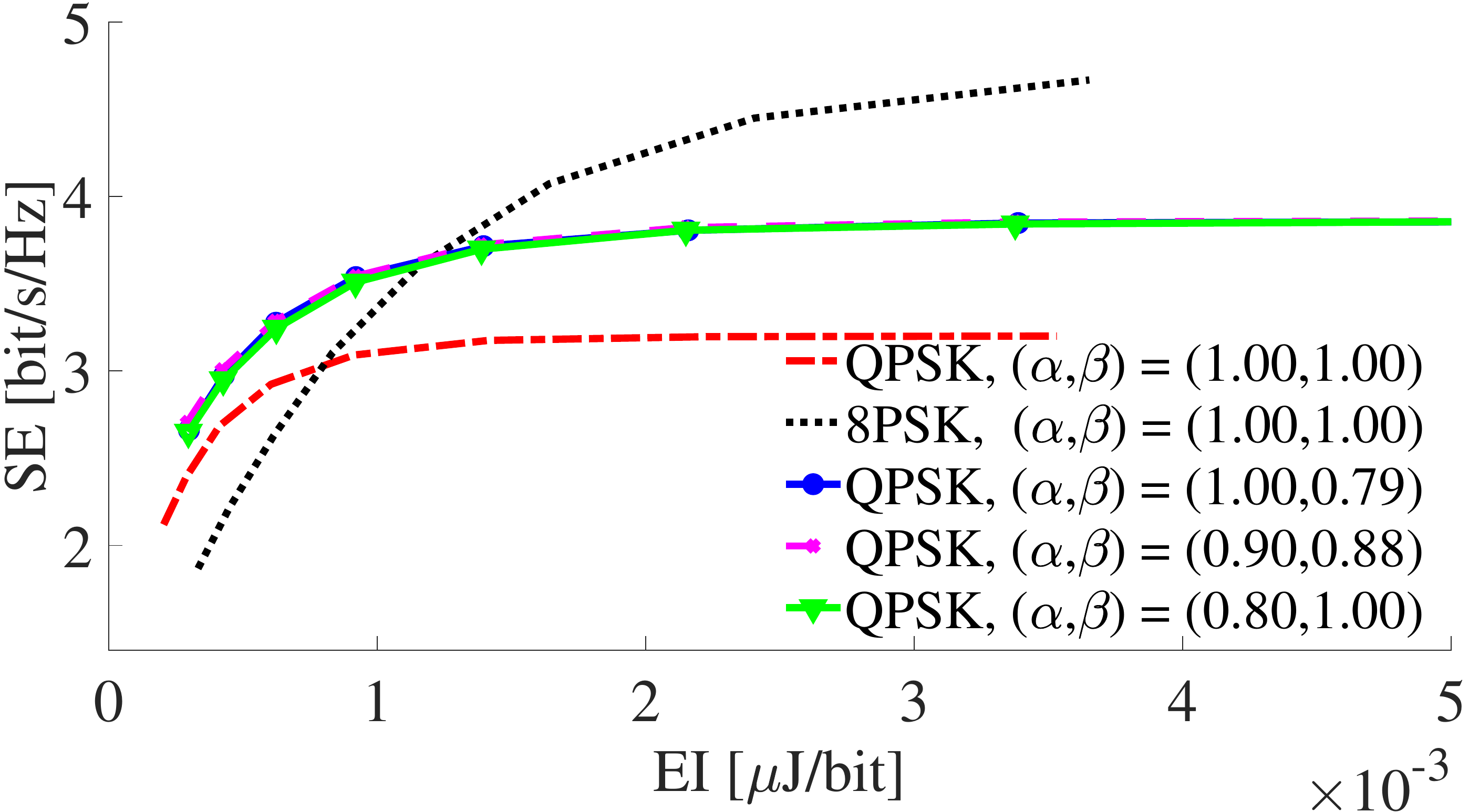}
}
\subfigure[\footnotesize Symbol error rate.]{%
\label{fig:Exp3-SER-gamma}%
\includegraphics[width=.45\linewidth]{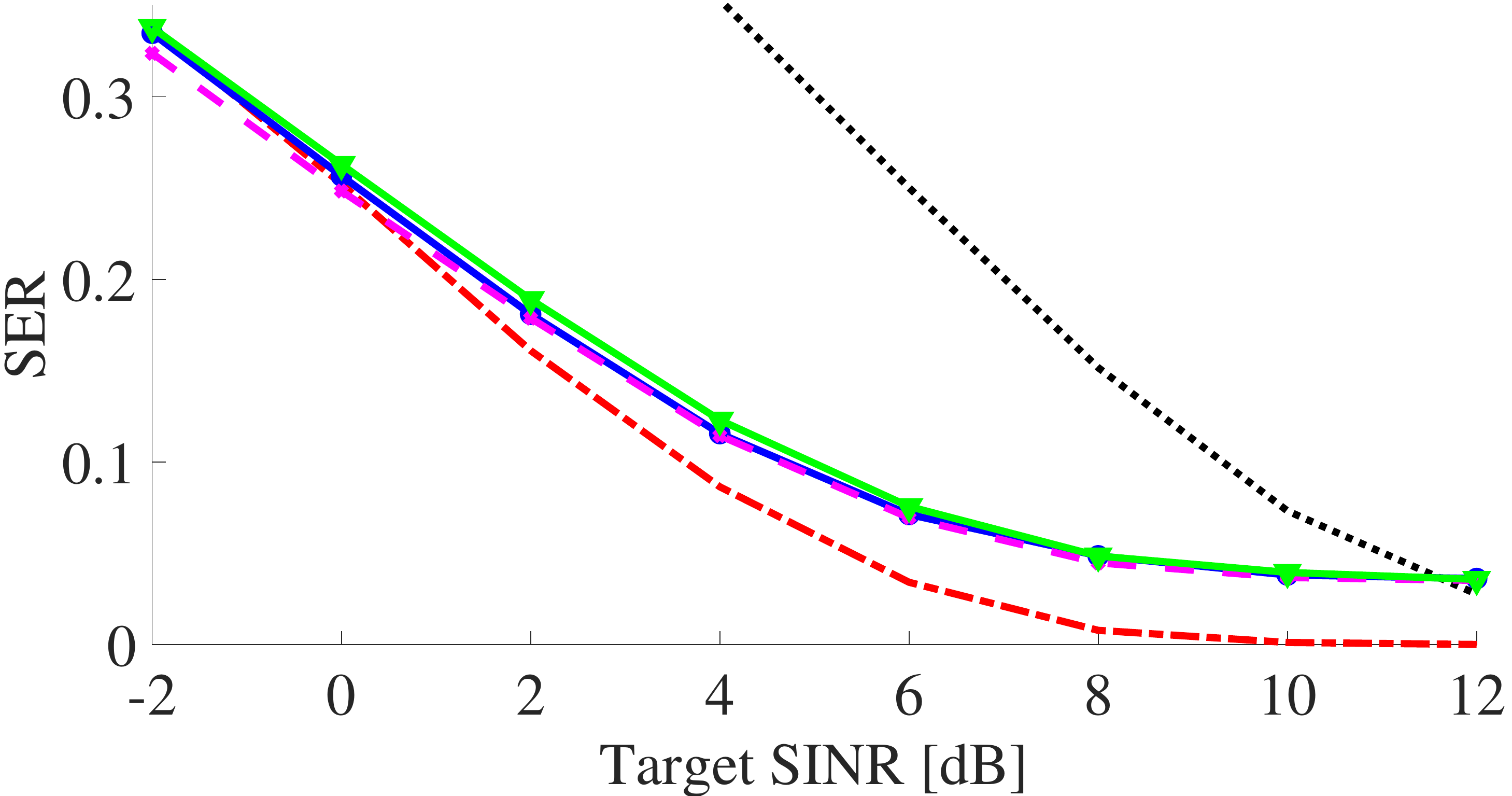}
}
\subfigure[\footnotesize Effective sum rate.]{%
\label{fig:Exp3-Rate-gamma}%
\includegraphics[width=.45\linewidth]{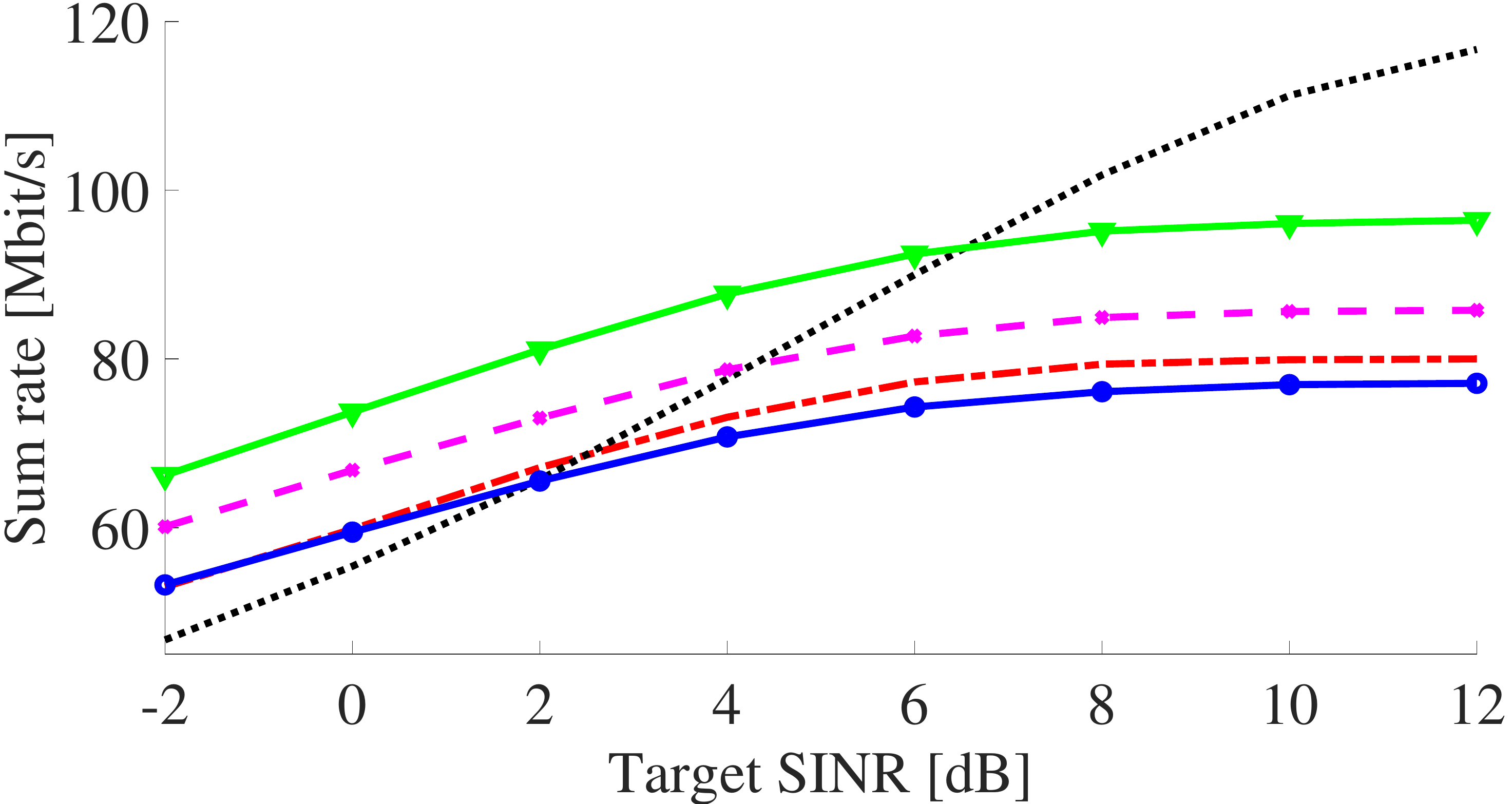}
}
\subfigure[\footnotesize Time-averaged total transmit power.]{%
\label{fig:Exp3-Power-gamma}%
\includegraphics[width=.45\linewidth]{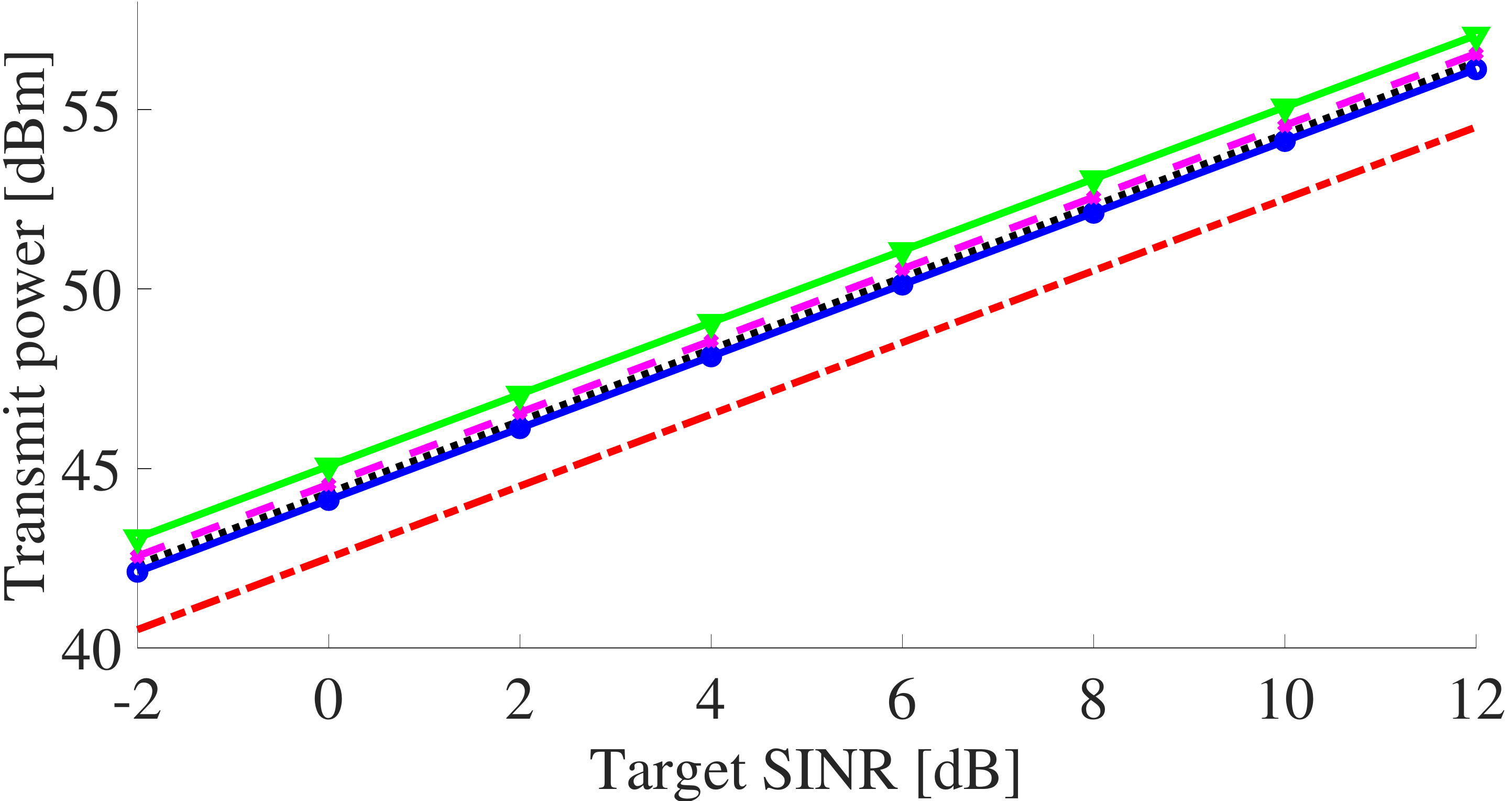}
}
\caption{(\emph{Experiment 2}) Performance comparison between different versions of the proposed space-time-frequency non-redundant SLP  across different target SINR values.  The legend in (a) applies to all figures.}\label{fig:exp2}
\end{figure*}


Fig.~\ref{fig:exp2} depicts the results. Fig~\ref{fig:Exp3-SEvsEI-gamma} shows that, as expected, if we jump from QPSK to 8PSK with no  frequency packed FTN signaling, i.e. with  $(\alpha,\beta) = (1,1)$, we can obtain higher spectral efficiency at the cost of a lower energy efficiency (i.e., higher energy inefficiency). There is a region in the SE~$\times$~EI plane, however,  where it is better to use QPSK rather than 8PSK. This happens for sufficiently small target SINR. On the other hand, when $\alpha < 1$ or $\beta < 1$, we find a region in the SE~$\times$~EI plane where it is better to use frequency packing and/or FTN signaling as compared to not using it. Again, this holds for sufficiently small target SINR, since one can get higher spectral efficiency with 8PSK without packing nor acceleration with the same energy efficiency of the other techniques for sufficiently large target SINRs. Fig~\ref{fig:Exp3-SEvsEI-gamma} also lets clear the equivalence among the different techniques with $\alpha < 1$ and/or $\beta < 1$ in the SE~$\times$~EI plane. The same virtual equivalence is observed in terms of SER performance in Fig.~\ref{fig:Exp3-SER-gamma}; but these techniques differ in other figures of merit, as explained in the following discussion. 

Fig.~\ref{fig:Exp3-Rate-gamma} shows that the effective sum rate of the accelerated schemes ($\alpha < 1$) is superior than the non-accelerated ones. It is interesting to note that the frequency-packed non-accelerated transmission with $(\alpha, \beta) = (1, 0.79)$ achieves virtually the same throughput of the QPSK with $(\alpha, \beta) = (1, 1)$ but using only $\xi_M(\beta) \approx 80\%$ of its bandwidth. Fig.~\ref{fig:Exp3-Power-gamma} shows that the technique that requires more transmit power is the one with $(\alpha, \beta) = (0.80, 1)$, which partially explains from where it comes its outstanding throughput performance. We therefore note, although the techniques with $\alpha < 1$ and/or $\beta < 1$ are virtually equivalent in the SE~$\times$~EI plane,  there is a trade-off among bandwidth usage, transmit power consumption, and throughput; this trade-off is controlled by the parameters $\alpha$ and $\beta$, which confer added flexibility to the system designer. 




\subsection{Experiment 3}



We now study the SLP performance for FTN signaling with and without frequency packing and our main goal is to verify the theoretical prediction in Proposition~\ref{prop:minAlpha}. We consider $M = 32$ subcarriers, $N = 4$ transmitting antennas, $K = N$ users, a QPSK constellation, and a target SINR $\gamma$ of 6~dB. Besides the baseline ZF linear precoder with full redundancy $R = \nu$, two space-time-frequency SLP precoders are employed according to the following guard-interval lengths: (i) $R = \left\lceil\frac{\delta}{4}\right\rceil$, called `Quarter-delay red. SLP' and (ii) $R = \left\lceil\frac{\delta}{2}\right\rceil$, called `Half-delay red. SLP'. In the simulations, we consider multipath-fading physical channels and the resulting effective channel order $\nu$ was around $25$, whereas the group delay $\delta$ was around $10$, depending on the specific values of $\alpha$ and $\beta$. We consider $\beta \in \{0.80, 0.90, 1\}$. According to Proposition~\ref{prop:minAlpha}, for each value of $\beta$, one has a specific value for $\alpha_{\min}$; in this case, $\alpha_{\min}\in \{0.99, 0.89, 0.80\}$. We therefore chose a range of values for $\alpha$ according to these minimum values.\footnote{Unfortunately, it was not possible to go much below the minimum value of $\alpha$ that guarantees information losslessness without facing some numerical issues, which limited the range of values that could be tested for the acceleration factor.}

\begin{figure*}[!t]
\centering
\subfigure[\footnotesize $\beta = 0.80$ and $\alpha_{\min} \approx 0.99$.]{
\label{fig:Exp2-SE-alpha-beta-80}
\includegraphics[width=.3\linewidth]{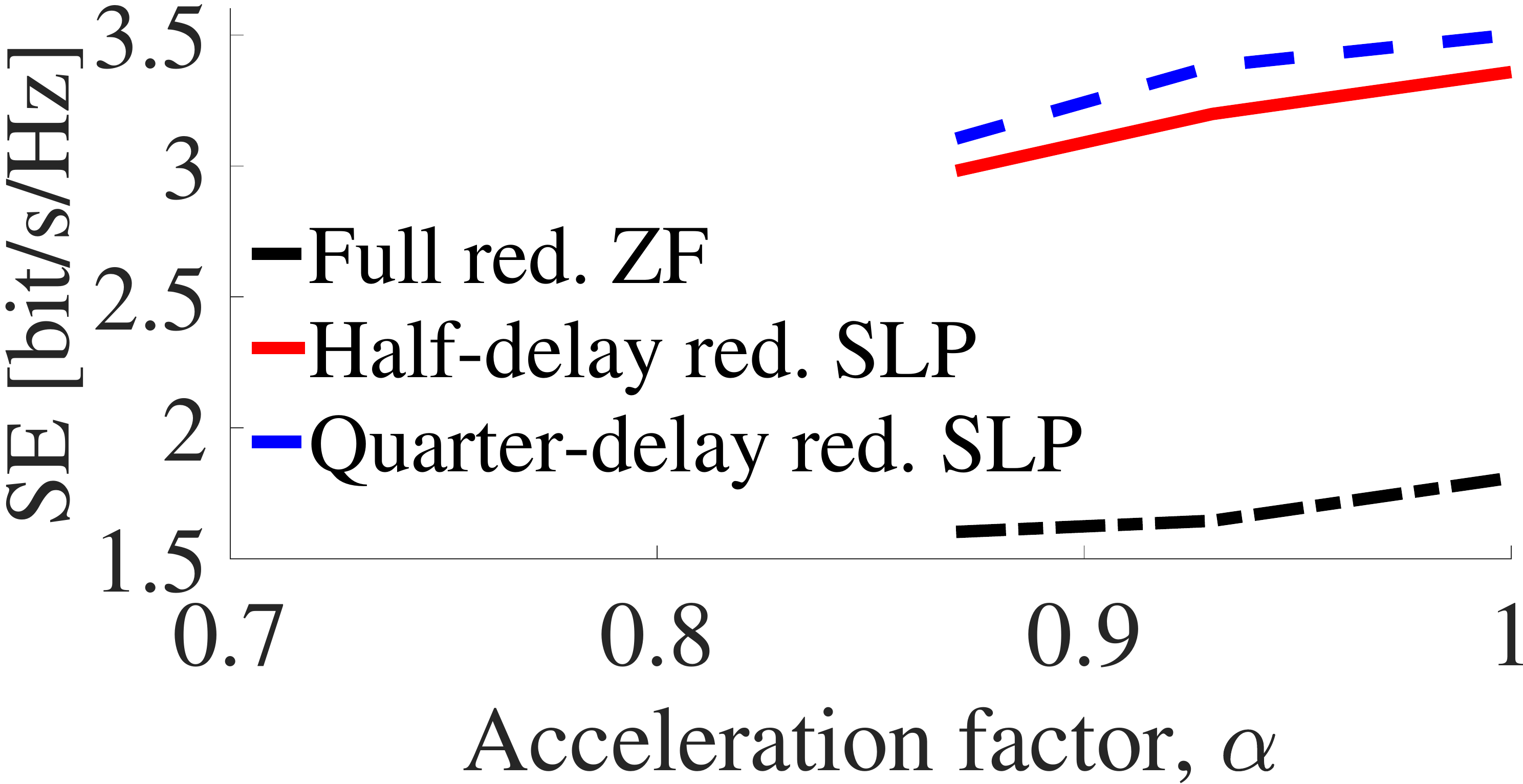}
}
\subfigure[\footnotesize $\beta = 0.90$ and $\alpha_{\min} \approx 0.89$.]{
\label{fig:Exp2-SE-alpha-beta-90}
\includegraphics[width=.3\linewidth]{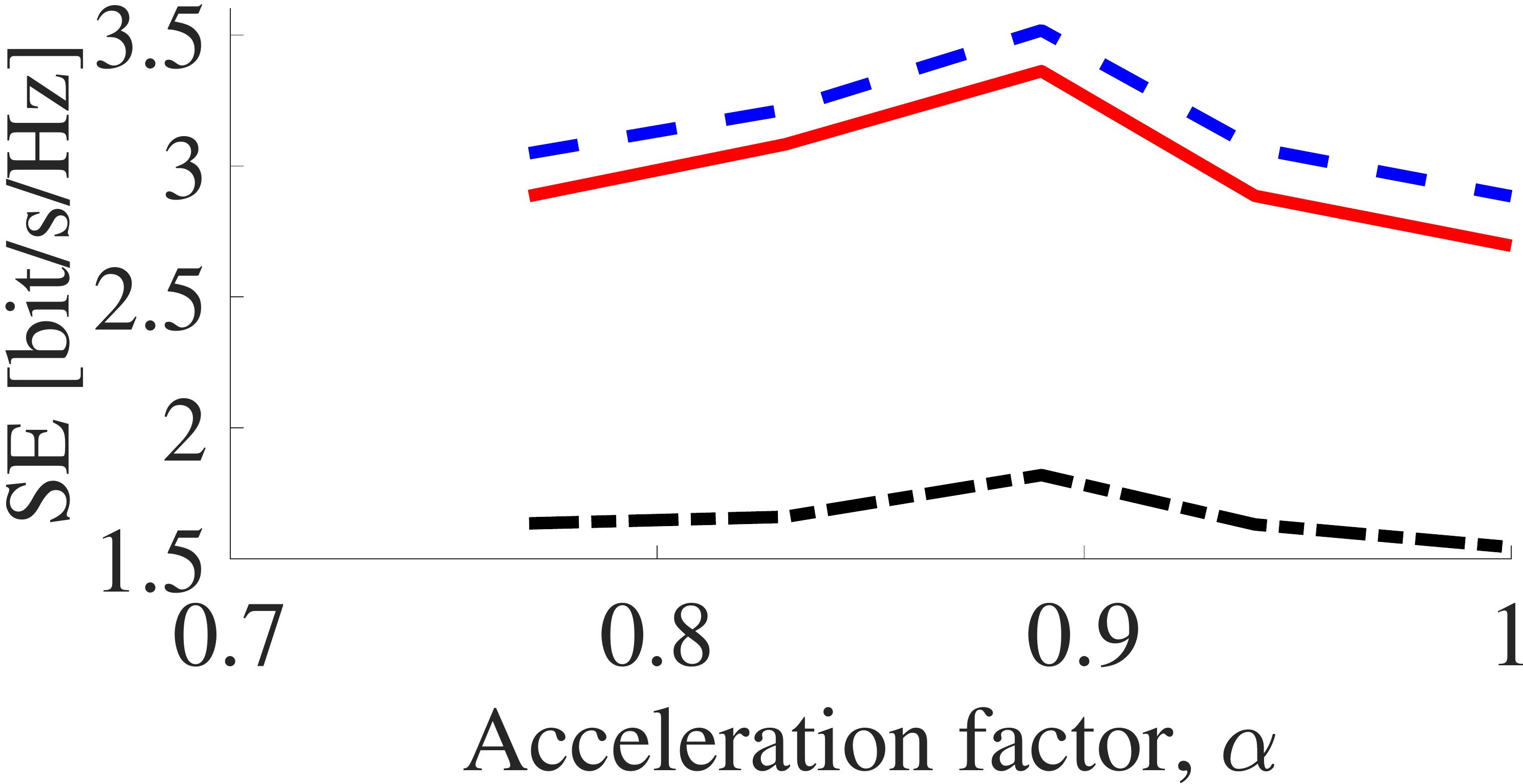}
}
\subfigure[\footnotesize $\beta = 1.00$ and $\alpha_{\min} = 0.80$.]{
\label{fig:Exp2-SE-alpha-beta-100}
\includegraphics[width=.3\linewidth]{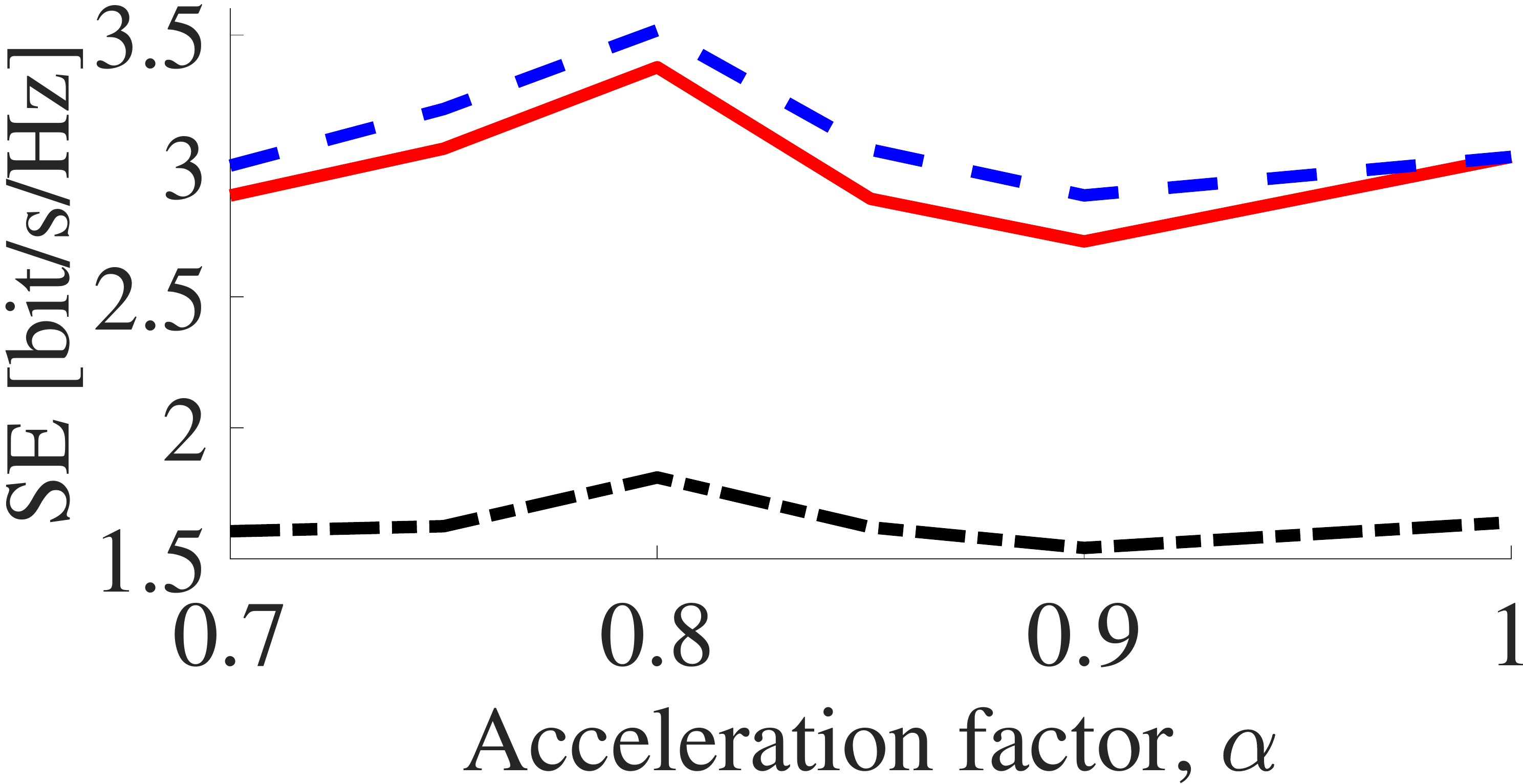}
}
\caption{(\emph{Experiment 3}) Spectral efficiency comparison between the ZF precoder with full redundancy and the proposed space-time-frequency SLPs with half-delay ($\left\lceil\delta/2\right\rceil$) and quarter-delay ($\left\lceil\delta/4\right\rceil$) redundancies, as function of $\alpha$ and for $\beta = 1$.  The legend in (a) applies to all figures.}\label{fig:exp3-se}
\end{figure*}

\begin{figure*}[!t]
\centering
\subfigure[\footnotesize $\beta = 0.80$ and $\alpha_{\min} \approx 0.99$.]{
\label{fig:Exp2-EI-alpha-beta-80}
\includegraphics[width=.3\linewidth]{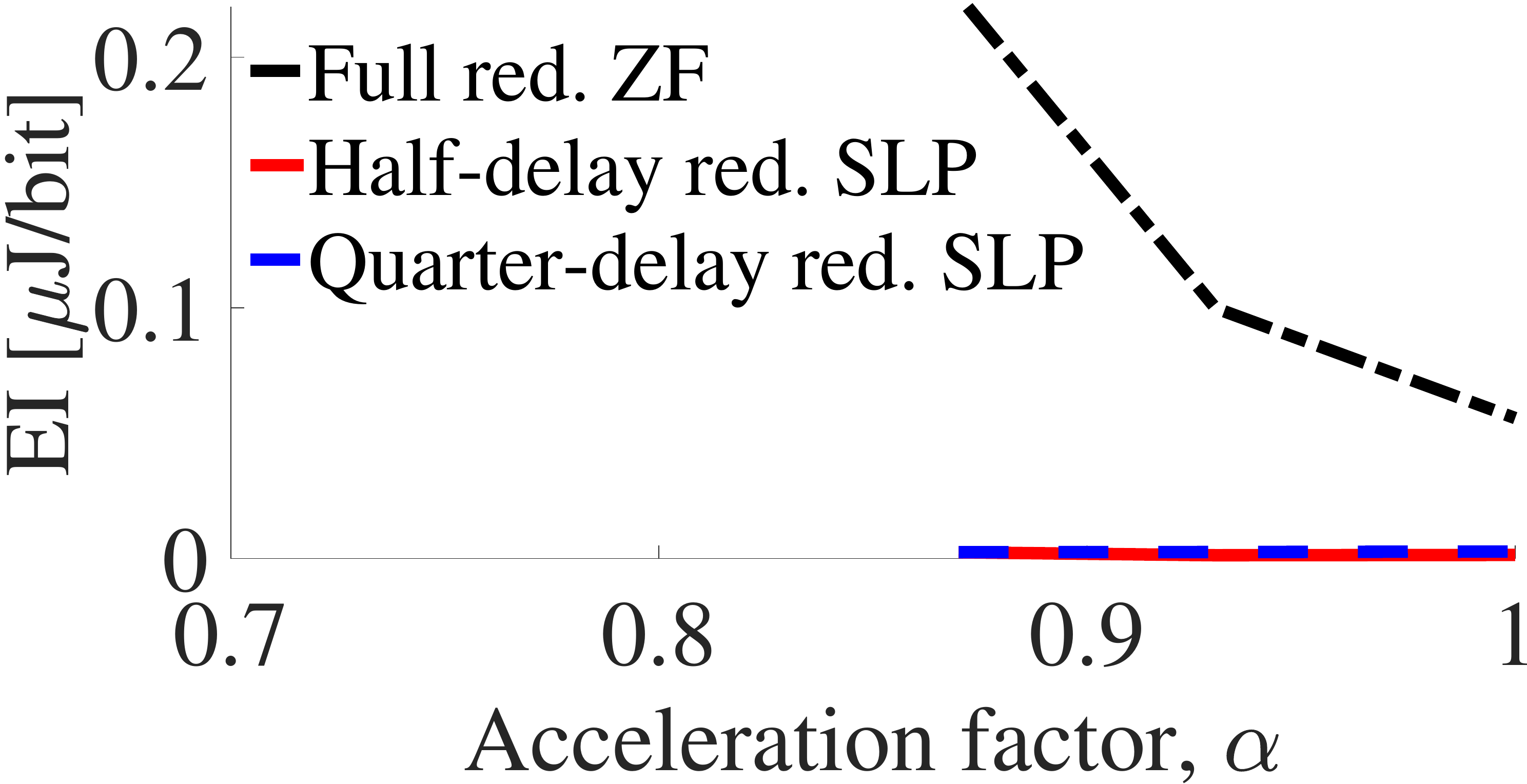}
}
\subfigure[\footnotesize $\beta = 0.90$ and $\alpha_{\min} \approx 0.89$.]{
\label{fig:Exp2-EI-alpha-beta-90}
\includegraphics[width=.3\linewidth]{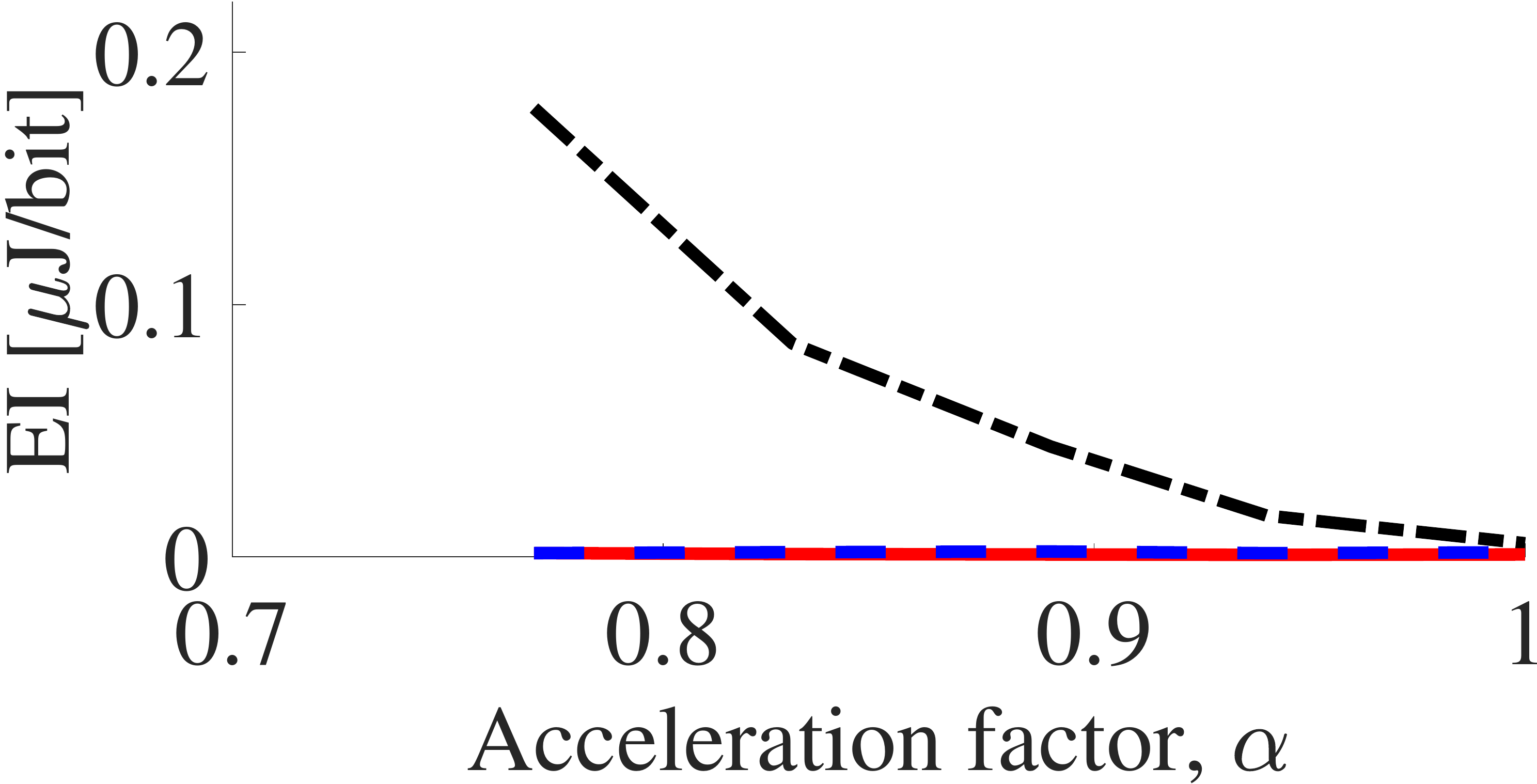}
}
\subfigure[\footnotesize $\beta = 1.00$ and $\alpha_{\min} = 0.80$.]{
\label{fig:Exp2-EI-alpha-beta-100}
\includegraphics[width=.3\linewidth]{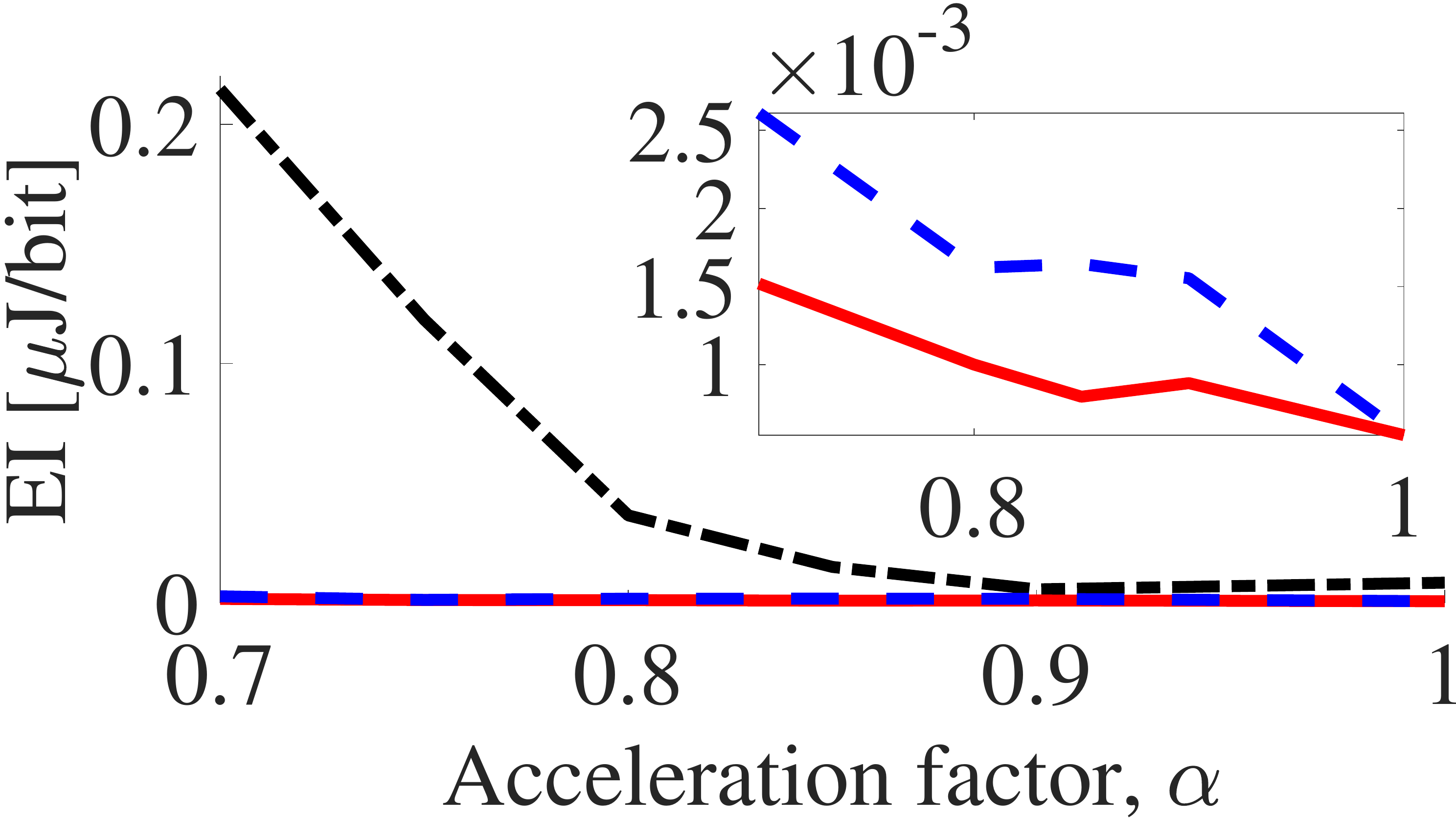}
}
\caption{(\emph{Experiment 3}) Energy inefficiency comparison between the ZF precoder with full redundancy and the proposed space-time-frequency SLPs with half-delay ($\left\lceil\delta/2\right\rceil$) and quarter-delay ($\left\lceil\delta/4\right\rceil$) redundancies, as function of $\alpha$ and for different values of $\beta$.  The legend in (a) applies to all figures.}\label{fig:exp3-ei}
\end{figure*}

Fig.~\ref{fig:exp3-se} depicts the spectral efficiency results. First of all, notice that the ZF precoder is quite inefficient since it is spending too much bandwidth resources with redundant signals (guard interval) that do not carry information. The proposed redundant precoder with $R = \left\lceil\delta/4\right\rceil$ attains the highest spectral efficiency among the tested schemes. The peak of the spectral efficiency is always achieved at the point $\alpha = \alpha_{\min}$. It is worth mentioning that such spectral efficiency enhancement of the proposed SLP when compared to the baseline scheme comes at the price of a higher computational burden.  

Fig.~\ref{fig:exp3-ei} shows the energy inefficiency results. The proposed redundant precoders are the most energy efficient ones, wherein the minimum-redundancy scheme with $R = \left\lceil\delta/2\right\rceil$ attains the highest energy efficiency (lowest inefficiency). Looking at the ZF curves, one can see that when trying to accelerate below the minimum value of $\alpha$ that guarantees information losslessness, the scheme starts to spend significantly more transmit power without having gains in spectral efficiency (see Fig.~\ref{fig:exp3-se}). Although less noticeable in this particular experiment, the same behavior is observed for the proposed non-redundant systems.

\section{Concluding Remarks}\label{sec:conc}
This paper addressed the problem of jointly handling multi-user, intersymbol, and intercarrier interference in downlink multi-antenna multicarrier transmissions through frequency-selective channels. More specifically, controlled intersymbol and intercarrier interference were introduced via frequency-packed faster-than-Nyquist signaling. In this context, redundant block-based space-time-frequency symbol-level precoding schemes were proposed. The introduction of redundancy in the block-based transmissions achieved a trade-off between constructive and destructive interference effects at the user terminals. A complete characterization of the interblock interference (IBI), taking into account that the resulting effective channel models can generate interference across many multicarrier symbols (not only two adjacent symbols), was provided considering the resulting group delay and its relationship with the minimum redundancy required for destructive IBI-free transmissions. Theoretical results showed the monotonicity of the error-free spectral efficiency with respect to the sampling time, as well as the minimum sampling time to guarantee information-losslessness transmissions as a function of the roll-off factor of the transmitting filter, the frequency-packing factor, and the number of subcarriers. Numerical results corroborated the theoretical predictions and showed that the proposed schemes can outperform zero-forcing precoders in terms of achieving a better balance between spectral and energy efficiencies. Future works include formulating and efficiently solving the problem of optimally setting the group delay of the effective channel model.


%
%
%

\appendices

\section{Proof of Proposition~\ref{prop:seGainsAlpha}}
\label{proof:prop:seGainsAlpha}
For fixed $\beta, b, r_{\rm c}, \rho, M$, as ${\rm SE}_0(\alpha,\beta) \propto \frac{M}{M+R_{\alpha,\beta}}\frac{1}{\alpha}$ and
\begin{align}
\frac{M}{M+R_{\alpha,\beta}}\frac{1}{\alpha} 
&=\frac{M}{\alpha M+\alpha \left\lfloor\frac{\alpha'}{\alpha}R_{\alpha',\beta}\right\rfloor}\nonumber\\
&\geq\frac{M}{\left(\frac{\alpha}{\alpha'}\right) M+ R_{\alpha',\beta}}\frac{1}{\alpha'}\nonumber\\
&=\frac{M}{M+R_{\alpha',\beta}}\frac{1}{\alpha'}\left(\frac{1}{1-\frac{\left[1-\left(\frac{\alpha}{\alpha'}\right)\right]M}{M+R_{\alpha',\beta}}}\right)\,,
\end{align}
then, when $\alpha < \alpha'$, one has $\frac{\left[1-\left(\frac{\alpha}{\alpha'}\right)\right]M}{M+R_{\alpha',\beta}} < 1$, thus implying that $\frac{M}{M+R_{\alpha,\beta}}\frac{1}{\alpha} > \frac{M}{M+R_{\alpha',\beta}}\frac{1}{\alpha'} \Leftrightarrow {\rm SE}_0(\alpha,\beta) > {\rm SE}_0(\alpha',\beta)$.

\section{Proof of Proposition~\ref{prop:minAlpha}}
\label{proof:prop:minAlpha}
From Lemma~\ref{lem:minAlpha}, we note that the necessary and sufficient condition for information losslessness  can be rewritten as $\tilde{F}^2({\rm j}\omega) > 0, \forall \omega \in \mathbb{R}$, for  $\tilde{F}^2({\rm j}\omega) \triangleq \sum_{i \in \mathbb{Z}}\left\vert F\left({\rm j}\frac{(\omega + 2\pi i)}{T_{\rm s}}\right)\right\vert^2$. As $f(t)$ is a square-root $T_{\beta}$-Nyquist filter with roll-off $\rho$, then $F\left({\rm j}\frac{(\omega + 2\pi i)}{T_{\rm s}}\right) = F\left(\frac{{\rm j}}{\alpha}\frac{(\omega + 2\pi i)}{T_1}\right) = 0$ for $\omega \not\in\left(-\frac{\alpha\xi_M(\beta)(1+\rho)\pi}{T_1}-\frac{2\pi i}{T_1}, \frac{\alpha\xi_M(\beta)(1+\rho)\pi}{T_1}-\frac{2\pi i}{T_1}\right)$. In this case, $\tilde{F}^2({\rm j}\omega) > 0, \forall \omega \in \mathbb{R}$ iff the intersection of the adjacent supports of $F\left({\rm j}\frac{(\omega + 2\pi i)}{T_{\rm s}}\right)$ and $F\left({\rm j}\frac{(\omega  + 2\pi (i+1))}{T_{\rm s}}\right)$ is nonempty; iff: $-\frac{\alpha\xi_M(\beta)(1+\rho)\pi}{T_1}-\frac{2\pi i}{T_1} \leq \frac{\alpha\xi_M(\beta)(1+\rho)\pi}{T_1}-\frac{2\pi (i+1)}{T_1} \Leftrightarrow \alpha\cdot\xi_M(\beta)\cdot(1+\rho) \geq 1$.

\section{Proof of Proposition~\ref{prop:noBlks}}
\label{proof:prop:noBlks}
Based on~\eqref{eq:delay} and~\eqref{subeq:HibiF-def}, the entries of the backward IBI matrices ($\boldsymbol{H}_{{\rm IBI}_{k,n}}^{\rm (b)}\![b]$) will be zero when $-bP  + p_{\rm r} - p_{\rm c} + q_{\delta}P + \rho_{\delta} < 0 \Leftrightarrow p_{\rm r} - p_{\rm c} < (b-q_{\delta})P-\rho_{\delta}$ for all pair $(p_{\rm r},p_{\rm c}) \in \mathscr{P}^2$. As $p_{\rm r} - p_{\rm c} \in \{-(P-1),\dots,(P-1)\}$ and $\rho_{\delta}\in \mathscr{P}$, the former inequality certainly holds when $b > q_{\delta}+1$, and it may hold when $b = q_{\delta}+1$ (whenever $\rho_{\delta} = 0$, i.e., whenever $\delta$ is a multiple of $P$). In other words, $\boldsymbol{H}_{{\rm IBI}_{k,n}}^{\rm (b)}\![b] = {\bf 0}_{P \times P}$ for $b > q_{\delta}+1$ or $(b,\rho_{\delta}) = (q_{\delta}+1,0)$, thus implying that $\mathscr{B}^{\rm (b)} = \{1,\dots,q_{\delta}+1\}$ in general. Similarly,  noticing that, based on~\eqref{eq:M-def}, one can write $\nu = (B-3) P + \rho_{\nu} + 1$, with $\rho_{\nu}$ being an integer number in the set $\mathscr{P}$, then it follows from~\eqref{subeq:HibiB-def} that the entries of the forward IBI matrices ($\boldsymbol{H}_{{\rm IBI}_{k,n}}^{\rm (f)}\![b]$) will be zero when $bP  + p_{\rm r} - p_{\rm c} + q_{\delta}P + \rho_{\delta} > (B-3) P + \rho_{\nu} + 1 \Leftrightarrow p_{\rm c} - p_{\rm r} < [b-(B-3-q_{\delta})]P-(1+\rho_{\nu}-\rho_{\delta})$ for all pair $(p_{\rm r},p_{\rm c}) \in \mathscr{P}^2$. As $p_{\rm c} - p_{\rm r} \in \{-(P-1),\dots,(P-1)\}$ and $1+\rho_{\delta}-\rho_{\nu}\in \{-(P-2),\dots,P\}$, the former inequality certainly holds when $b > B-2-q_{\delta}$, and it may hold when $b = B-2-q_{\delta}$ (whenever $\rho_{\delta} > \rho_{\nu}$; in this case, $\delta$ is not a multiple of $P$). In other words, $\boldsymbol{H}_{{\rm IBI}_{k,n}}^{\rm (f)}\![b] = {\bf 0}_{P \times P}$ for $b > B-2-q_{\delta}$, or $b = B-2-q_{\delta}$ and $\rho_{\delta} > \rho_{\nu} = \nu-(B-3)P-1$, thus implying that $\mathscr{B}^{\rm (f)} = \{1,\dots,B-2-q_{\delta}\}$ in general. Note that, from~\eqref{eq:delay}, the group delay satisfies $\delta <\nu$, which implies $(B-2-q_{\delta})P > P-(1+\rho_{\nu}-\rho_{\delta}) \geq 0$, thus yielding $B-2-q_{\delta} \geq 1$, so that the set $\mathscr{B}^{\rm (f)}$ is well-defined.   From~\eqref{eq:y-ISI-IBI}, the number of data blocks contributing to the $\ell^{\rm th}$ received block after sampling, synchronization, and buffering is up to $1+\left\vert\mathscr{B}^{\rm (f)}\right\vert+\left\vert\mathscr{B}^{\rm (b)}\right\vert = B$ whenever $\delta$ is such that $\rho_{\delta} \in \{1, 2, \ldots , \rho_{\nu}\}$, or otherwise (i.e., $\delta = q_{\delta}P + \rho_{\delta}$, with $q_{\delta}\in\mathbb{N}$ and $\rho_{\delta} \in\mathscr{P}\setminus \{1, 2, \ldots , \rho_{\nu}\}$) up to $B-1$.

\section{Proof of Proposition~\ref{prop:minR}}
\label{proof:prop:minR}
Firstly, note that $\boldsymbol{R}\boldsymbol{H}_{{\rm IBI}_{k,n}}^{\rm (b)}\![b]\boldsymbol{A}$ is an $M\times M$ matrix comprised of the entries $\left[ \boldsymbol{H}_{{\rm IBI}_{k,n}}^{\rm (b)}\![b] \right]_{p_{\rm r},p_{\rm c}}$, with $p_{\rm r} \in\{0,\cdots,M-1\}$ and $p_{\rm c} \in\{P-M,\cdots,P-1\}$. From~\eqref{subeq:HibiF-def}, backward-IBI-free transmissions are possible when $-bP + p_{\rm r}-p_{\rm c}+\delta < 0$ for all $b\in\mathscr{B}^{\rm (b)}$, $p_{\rm r} \in\{0,\cdots,M-1\}$,  and $p_{\rm c} \in\{P-M,\cdots,P-1\}$. The maximum value that the L.H.S. of the former inequality can assume occurs when $(b,p_{\rm r},p_{\rm c}) = (1,M-1,M-P)$. Thus, recalling that $P = M + R$, one must have $-(M + R) +(M-1) - R +\delta < 0 \Leftrightarrow 2R > \delta-1$, from which~\eqref{eq:minR} follows.

\balance
\bibliographystyle{IEEEtran}
\bibliography{refs.bib}

\end{document}